\documentclass[a4paper,11pt ]{article}
\usepackage{amsmath,mathtools,amsfonts,amssymb}
\usepackage{graphicx}
\usepackage[linktocpage=true]{hyperref}
\usepackage{a4wide}
\usepackage{caption}
\usepackage{subcaption}
\usepackage[usenames,dvipsnames]{color}

 \usepackage{tikz}
\usepackage{pgfplots}
\usetikzlibrary{calc}
\usetikzlibrary{shapes,decorations}
\usetikzlibrary{arrows}

\newcommand{\nn}{\nonumber}
\renewcommand{\d}{\mathrm{d}}
\newcommand{\lc}{\left(}
\newcommand{\rc}{\right)}

\newlength{\eqboxstorage}

\title{Holographic Construction of Excited CFT States}
\author{Ariana Christodoulou and Kostas Skenderis \\ \\
STAG Research Centre and Mathematical Sciences, \\
University of Southampton, UK \\
Email:  misc1g13@soton.ac.uk, K.Skenderis@soton.ac.uk}
\date{}

\begin{document}
\maketitle
\abstract{We present a systematic construction of bulk solutions that are dual to CFT excited states. The bulk solution is constructed 
perturbatively in bulk fields. The linearised solution is universal and depends only on the conformal dimension of the primary operator that is associated with the state via the operator-state correspondence, while higher order terms depend on detailed properties of the operator, such as its OPE with itself and generally involve many bulk fields. We illustrate the discussion with the holographic construction of the universal part of the solution for states of two dimensional CFTs, either on $R \times S^1$ or on $R^{1,1}$.  We compute the 1-point function both in the CFT and in the bulk, finding exact agreement. We comment on the relation with other reconstruction approaches.}
\newpage
\tableofcontents
\newpage

\section{Introduction}

A central question in holography is how the bulk is reconstructed from QFT data. In this paper we would like to ask and answer  a simpler question:``what is the bulk dual of a CFT state?'' While it has been clear since the early days of AdS/CFT that normalisable bulk solutions are related to states \cite{Balasubramanian:1998de},  a precise construction of a bulk solution given a state has not been available prior to this work\footnote{ 
A related question that received more attention over the years is the converse:  given a bulk solution with normalisable asymptotics what is the dual state? For such solutions, the leading order asymptotic behavior of the solution is related with the 1-point function of the gauge invariant operators in a state and from the 1-point functions one may extract information about the dual states. Examples of such computations include the computation of 1-point functions 
for the solutions corresponding to the Coulomb branch of ${\cal N}=4$ SYM  \cite{Skenderis:2006di}, the 1-point functions for the LLM solutions \cite{Lin:2004nb}  in \cite{Skenderis:2007yb} and 1-point functions for fuzzball solutions \cite{Kanitscheider:2006zf,Kanitscheider:2007wq, Giusto:2015dfa}.}.

The construction is an application of the real-time gauge/gravity dictionary \cite{Skenderis:2008dh, Skenderis2009} and it can be applied to any state that has a (super)gravity description. We will however focus on a simple example: a state that to leading order 
in a large $N$ limit can be described by a scalar field in a fixed AdS background. An additional motivation for studying this example is that the bulk solution appeared also in related work \cite{Hamilton:2006az} and we will discuss similarities and differences with that work.

Let us briefly review what is known about bulk reconstruction using the example of a scalar field in a fixed background, starting first with the case of Euclidean signature.
It is well known that a scalar field $\Phi$ of mass $m^2 = \Delta (\Delta -d)$ in AdS$_{d+1}$ is dual to an operator $\mathcal{O}_\Delta$ of dimension $\Delta$.
The bulk field has an asymptotic expansion of the form~\cite{DeHaro2000} 
\begin{equation} \label{as_exp}
\Phi(r,x) = r^{d-\Delta} \phi_{(0)}(x) + \dots + r^\Delta \log r^2 \psi_{(2\Delta - d)} (x) + r^\Delta \phi_{(2\Delta - d)} (x) + \dots
\end{equation}
where $r$ is the holographic (radial) direction and $x$ denotes the collective set of boundary coordinates.
$\phi_{(0)}(x)$ is the source for the dual operator and $\phi_{(2\Delta - d)}(x)$ is related to the 1-point function,
\begin{equation} \label{vev}
\langle \mathcal{O}_\Delta \rangle = (2\Delta -d) \phi_{(2\Delta -d)} (x) + X(\phi_{(0)}),
\end{equation}
where $X(\phi_{(0)})$ is a local function of the source $\phi_{(0)}$ (whose exact form depends on the bulk theory under discussion).
$\phi_{(0)}(x)$ and $\phi_{(2\Delta -d)}(x)$ are the only two arbitrary coefficient functions in the above expansion. 
All subleading terms down to $r^\Delta$ (including $\psi_{(2\Delta - d)}$ but not  $\phi_{(2\Delta -d)}(x)$) are locally related to $\phi_{(0)}(x)$ and similarly all terms that appear at higher orders can be determined in terms of $\phi_{(0)}$ and $\phi_{(2\Delta -d)}(x)$.  Thus, given the pair ($\phi_{(0)}(x)$, $\phi_{(2\Delta -d)}(x)$) one can iteratively construct a unique bulk solution. A different (non-perturbative)
argument for uniqueness is to note that the 1-point function is the canonical momentum $\pi_{\Delta}$ in a radial Hamiltonian formalism \cite{Papadimitriou:2004ap} and by a standard Hamiltonian argument, specifying a conjugate pair $(\phi_{(0)}, \pi_{\Delta})$ uniquely picks a solution of the theory. This argument however does not tell us whether the solution is regular in the interior.
Indeed in quantum field theory, the vacuum structure is a dynamical question: in general one cannot tune the value of $\langle \mathcal{O}_\Delta \rangle$. The counterpart of this statement is that a generic pair $(\phi_{(0)}, \pi_{\Delta})$ leads to a singular solution\footnote{Some of these pairs do not correspond to QFT data at all
while others are singular in supergravity but they would be regular in string theory.  It is not currently known how to distinguish between the two cases.} and it is regularity in the interior that selects $\langle \mathcal{O}_\Delta \rangle$.  

In Lorentzian signature new complications arise.  In the bulk, boundary conditions alone do not determine a unique solution:
Lorentzian AdS is a non--hyperbolic manifold. Indeed, there exist
normalisable modes which are regular in the interior and vanish at the boundary, leaving the boundary data unaffected.

On the QFT side, there are related issues. While in Euclidean signature there is only one type of correlator, in Lorentzian signature, there are multiple types of correlators (time-ordered, Wightman functions, advanced, retarded, etc.). In addition, one may wish to consider these correlators on non--trivial states (such as thermal states, states that spontaneously break some symmetries, general non-equilibrium states). All of this data may be nicely encoded by providing a contour in the complex time plane and considering the path integral defined along this  contour. Different types of correlators and different initial/final states are encoded by operator insertions along this contour.
This is known as the Schwinger-Keldysh formalism  \cite{Schwinger:1960qe, Bakshi:1962dv, Bakshi:1963bn, Keldysh:1964ud}.

A bulk version of this formalism was developed in~\cite{Skenderis:2008dh, Skenderis2009}: the gauge/gravity duality acts in a piece-wise fashion on the various parts of the time contour and appropriate matching condition are imposed at the corners. More specifically,
real time pieces of the contour are associated with Lorentzian AdS manifolds, imaginary time pieces with Euclidean AdS manifolds and the matching conditions require that 
the fields and their conjugate momenta are continuous across the different manifolds. In this way, the initial conditions are traded for  boundary condition in the Euclidean parts of the spacetime. In this formalism, imposing boundary conditions on the entire bulk manifold, uniquely specifies the bulk solution, as in the Euclidean case.

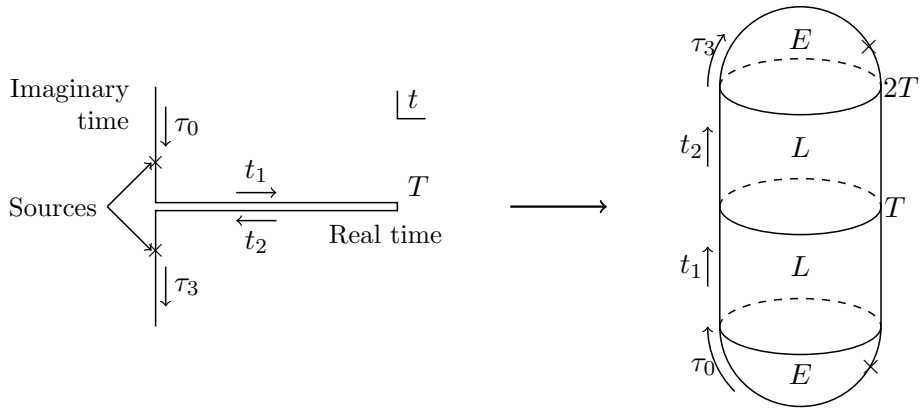
\begin{figure}[t]
\centering
\begin{tikzpicture}[line width = 0.2mm, scale=0.53]
\draw (0,3) -- (0,0.1) -- (6, 0.1) -- (6,-0.1) --(0,-0.1) -- (0, -3);
\node [anchor=east] at (0,2.5)  {\small \begin{tabular}{r} Imaginary \\ time \end{tabular} };
\node [anchor= west] at (0.2, 2) {$\tau_0$};
\node [anchor = west] at (0.2cm, -2 cm) {$\tau_3$};
\node [anchor = south] at (2.5 cm, 0.35 cm) {$t_1$};
\node [anchor = north] at (2.5 cm, -0.35 cm) {$t_2$};
\node [anchor =south west] at (6 cm,0 cm) {$T$};
\draw[->] (0.25, 2.5) -- (0.25,1.5);
\draw[->] (0.25, -1.5) -- (0.25, -2.5);
\draw[line width =0.2 mm, ->] (2 cm,0.35 cm) -- (3 cm, 0.35 cm); 
\draw[line width =0.2 mm, ->] (3 cm,-0.35 cm) -- (2 cm, -0.35 cm); 
\node [anchor=center] at ( 0cm,1.1 cm) {\small$\times$};
\node [anchor=center] at (0cm, -1.1 cm) {\small $\times$};
\draw[line width=0.2mm, ->] (-1.2,0) -- (-0.1,1.1);
\draw[line width=0.2mm, ->] (-1.2,0) -- (-0.1,-1.1);
\node [anchor = east] at (-1.2,0) {\small Sources};
\node [anchor = north] at (5.7cm, -0.2cm) {\small Real time};
\draw (6cm, 2.9cm) --(6cm, 2.2cm )--(6.7cm,2.2 );
\node [anchor = south west] at (6, 2.2) {$t$};
\draw [line width = 0.3mm, ->] (8.8, 0) -- (11.2,0);
\draw[line width=0.2 mm ,dashed] (14 cm ,3 cm) arc (-180:-360:2 cm and 0.7 cm);
\draw[line width=0.2 mm] (14 cm ,3 cm) arc (180:360:2 cm and 0.7 cm);
\draw[line width = 0.2mm] (14 cm, 3 cm ) -- (14 cm, -3 cm);
\draw[line width=0.2 mm ,dashed] (14 cm ,0 cm) arc (-180:-360:2 cm and 0.7 cm);
\draw[line width=0.2 mm] (14 cm ,0 cm) arc (180:360:2 cm and 0.7 cm);
\draw[line width = 0.2mm] (18 cm, 3 cm ) -- (18 cm, -3cm);
\draw[line width=0.2 mm ,dashed] (14 cm , -3 cm) arc (-180:-360:2 cm and 0.7 cm);
\draw[line width=0.2 mm] (14 cm , -3 cm) arc (180:360:2 cm and 0.7 cm);
\draw[line width=0.2mm] (14 cm, 3 cm) arc (180:0:2 cm);
\draw[line width=0.2mm, ->] (13.7 cm, 3 cm) arc (180:145:2.3 cm);
\node at (13.6 cm, 4 cm) {$\tau_3$};
\draw[line width=0.2mm, ->] (13.7 cm, 1 cm) -- (13.7, 2 cm);
\node at (13.3 cm, -1.5 cm) {$t_1$};
\node at (16 cm, 1.5 cm) {$L$};
\draw[line width = 0.2mm, ->] (13.7 cm, -2 cm) -- (13.7 cm, -1 cm);
\node at (13.3 cm, 1.5 cm) {$t_2$}; 
\node [anchor = center] at (16 cm, -1.5 cm) {$L$};
\draw[line width=0.2mm] (14 cm, -3 cm) arc (180:360:2 cm);
\node at (18.35 cm, 0 cm) {$T$};
\node at (18.5 cm, 3 cm) {$2T$};
\draw[line width=0.2mm, <-] (13.7 cm, -3 cm) arc (180:225:2.3 cm);
\node at (13.6 cm, -4 cm) {$\tau_0$};
\node [anchor= center] at (16 cm, 4.2 cm) {$E$};
\node at (16 cm, -4.2 cm) {$E$};
\node at (17.7 cm, 4 cm) {$\times$};
\node at (17.75 cm, -4 cm) {$\times$};
\end{tikzpicture}
\caption{In--in time contour (left) and corresponding AdS manifold (right). The manifolds labeled by $L$ are empty Lorentzian AdS and those labeled by $E$ are empty, Euclidean AdS.}
\label{time contour and manifold}
\end{figure}

This is a general method that may be used to study correlation functions in general non-equilibrium states.  In this paper we will use it to
construct a bulk solution that corresponds to an excited CFT state.
By the operator-state correspondence any such state may be obtained by acting with scalar primary operators 
$\mathcal{O}_{\Delta}$ on the CFT vacuum, 
\begin{equation}
|\Delta \rangle = \mathcal{O}_{\Delta} |0\rangle.
\end{equation}
In the Schwinger-Keldysh formalism, in-in correlators in this state may be obtained by considering the in-in contour $\mathcal{C}$
on the left panel of 
Fig. \ref{time contour and manifold}. On the gravity side we consider the manifold corresponding to the in--in field theory time contour shown  in the right panel of figure \ref{time contour and manifold}. The operator $\mathcal{O}_{\Delta}$  corresponds to a massive bulk scalar field
and we will  solve the scalar field equation in all four parts of the bulk spacetime.
The boundary conditions we use are sources turned on in the two Euclidean manifolds , {\it i.e.}  $\phi_{(0)}(x)\neq 0$ for $x \in \partial E$ where $\partial E$ the boundary of the Euclidean manifolds.
In the Lorentzian manifolds we want purely normalisable solutions so we set the sources equal to zero, {\it i.e.} $\phi_{(0)}(x) =0$ for $x \in \partial L$ where $\partial L$ is the boundary of the Lorentzian manifolds.

This paper is organised as follows. In the next section we discuss the QFT computation of the expectation value of operators in this state.
We will later compute the same quantity by a bulk computation in order to confirm that the bulk solution indeed represents the state
it should. In section \ref{sec:global} we discuss the construction of the solution dual to a state of a two dimensional CFT on $R \times S^1$, while in section \ref{sec:poincare} we solve the same problem for a CFT on $R^{1,1}$. We conclude in section \ref{sec:discussion}, where we also discuss the relation with the work \cite{Hamilton:2006az}.  Appendix \ref{sec:appendix} contains a number of technical details relevant for section  \ref{sec:poincare}.

As this paper was finalised, we received \cite{Botta-Cantcheff:2015sav} which presents related material. 
Preliminary version of this work was presented in a number of international workshops \cite{conferences}.

\section{Quantum field theory considerations} \label{QFT}

In this section we setup the problem using the Schwinger-Keldysh formalism.
Let us denote by $\phi_{(0)}$ the source that couples to $\mathcal{O}_{\Delta}$.
We would like to compute expectation values in the  state $|\Delta \rangle = \mathcal{O}_{\Delta} |0\rangle$, inserted 
at $\vec{x}=t=0$. To realise this set up we consider the contour shown in Fig. \ref{fig:in-in_time_contour}. 
We  insert the operator $\mathcal{O}_{\Delta}$ at small imaginary distance $\tau_0 =- \epsilon$ at $t=0$  and at $\tau_3=\epsilon$ at $t_2=2T$, where $\tau_0, t_1, t_2$ and $\tau_3$  are contour times in the four segments. In complexified time the insertions are 
at $t = 0 + i \epsilon$ and $t=0-i \epsilon$. Performing the Euclidean path integral over the imaginary part of the contour 
provides the initial and final conditions for the Lorentzian path integral.
Altogether the path integral under consideration is  
\begin{equation} \label{partfun}
Z\left[\phi_{(0)};   \mathcal{C}\right] =   \int \left[{\mathcal D} \phi\right]  \exp  \left[-i \int_\mathcal{C}\d t \d^{d-1}x \sqrt{-g_{(0)}} 
\left({\mathcal L}_{QFT}  
+\phi_{(0)}(x)\mathcal{O}_\Delta(x)\right)  \right]
\end{equation}
If we compute this path integral for general $\phi_{(0)}(x)$ and then differentiate w.r.t. $\phi_{(0)}^+$ and  $\phi_{(0)}^-$,
where $\phi_{(0)}^\pm = \phi_{(0)}(0_\pm, \vec{0})$ and $0_\pm = 0\pm i \epsilon$, and then set to zero the sources in the imaginary part of the contour, the resulting expression will be the desired  generating functional of in-in correlators in the state $|\Delta \rangle$. 

In later sections we will construct the gauge/gravity analogue of (\ref{partfun}).  Corresponding to $\phi_{(0)}$ there is bulk scalar field  $\Phi$ and 
the best we can currently do  holographically is to construct  (\ref{partfun}) perturbatively in the bulk fields (or perturbatively in a large $N$ limit, see below). Correspondingly we will consider the source $\phi_{(0)}(x)$ 
in the imaginary part as being infinitesimal, with the product of the two sources at the same point set to zero, $(\phi_{(0)}(x))^2=0$, so that we generate a single insertion. If we relax this condition we will generate states that are superpositions of the states  associated with ``single trace'' and ``multi-trace'' operators. The path integral  (\ref{partfun}) with $\phi_{(0)}(x)$ infinitesimal also contains
terms linear in the sources which would not contribute if we were to 
differentiate w.r.t. both $\phi_{(0)}^+$ and  $\phi_{(0)}^-$. However, these linear terms still  provide a non-trivial check that we are constructing holographically the correct path integral and as such we will consider them in detail.

\begin{center}
\begin{figure}[t]
\centering
\begin{tikzpicture}[line width = 0.2mm, scale=0.6]
\draw (0,3) -- (0,0.1) -- (6, 0.1) -- (6,-0.1) --(0,-0.1) -- (0, -3);
\node [anchor= west] at (0.2, 2) {$\tau_0$};
\node [anchor = west] at (0.2cm, -2 cm) {$\tau_3$};
\node [anchor = south] at (3 cm, 0.35 cm) {$t_1$};
\node [anchor = north] at (3 cm, -0.35 cm) {$t_2$};
\node [anchor =south west] at (6 cm,0 cm) {$T$};
\draw[->] (0.25, 2.5) -- (0.25,1.5);
\draw[->] (0.25, -1.5) -- (0.25, -2.5);
\draw[line width =0.2 mm, ->] (2.5 cm,0.35 cm) -- (3.5 cm, 0.35 cm); 
\draw[line width =0.2 mm, ->] (3.5 cm,-0.35 cm) -- (2.5 cm, -0.35 cm); 
\node [anchor=center] at ( 0cm,1.1 cm) {\small$\times$};
\node [anchor=center] at (0cm, -1.1 cm) {\small $\times$};
\node [anchor = east] at (0 cm, 1.1 cm) {$\epsilon$};
\node [anchor = east] at (0 cm, -1.1 cm) {$-\epsilon$};
\end{tikzpicture}
\caption{In--in complex time contour with operator insertions at $t = 0 \pm i \epsilon$.}
\label{fig:in-in_time_contour}
\end{figure}
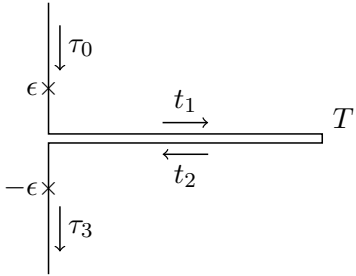
\end{center}

Let ${\cal O}_i$ be gauge invariant operators. Their 1-point function is given by
\begin{align}
\langle \mathcal{O}_i (t,\vec{x}) \rangle 
&=\int \left[{\mathcal D} \phi \right]  \mathcal{O}_i(t,\vec{x})\exp \bigg[-i \int_\mathcal{C} \d t' \d^{d-1}\vec{x}^\prime \sqrt{-g_{(0)}}  \left({\mathcal L}_{QFT}  \right. \nonumber \\
& \left. \phantom{\int \left[{\mathcal D} \phi \right]  \mathcal{O}_{\Delta}(x)\exp \bigg[} 
\qquad +\phi_{(0)}(x^\prime) \mathcal{O}_\Delta(x^\prime)\right) 
\bigg].
\end{align}
Expanding in the sources we obtain
\begin{align}
\langle \mathcal{O}_i (t,\vec{x})  \rangle=& \phi_{(0)}^+  \langle 0 | \mathcal{O}_\Delta (0_{+}, \vec{0})   \mathcal{O}_i  (t,\vec{x}) |0\rangle + \phi_{(0)}^-
 \langle 0| \mathcal{O}_i  (t,\vec{x}) \mathcal{O}_\Delta (0_{-}, \vec{0})  |0\rangle \nonumber \\
& \qquad+\phi_{(0)}^+  \phi_{(0)}^-  \langle 0 | \mathcal{O}_\Delta (0_+, \vec{0})  \mathcal{O}_i(t,\vec{x})  \mathcal{O}_\Delta (0_-, \vec{0})  | 0 \rangle.  \label{QFT:1pt} \\
=& \phi_{(0)}^+  \langle \Delta|  \mathcal{O}_i  (t,\vec{x}) |0\rangle + \phi_{(0)}^- \langle 0| \mathcal{O}_i  (t,\vec{x})| \Delta\rangle +\phi_{(0)}^+  \phi_{(0)}^-  \langle \Delta|  \mathcal{O}_i  (t,\vec{x}) | \Delta\rangle \nonumber 
\end{align} 
Note that the correlators that appear here are all Wightman functions, as can be seen from the time contour.  The expectation value of ${\cal O}_i$ in the state $|\Delta \rangle$ appears in the terms quadratic
in the sources. As mentioned above, we kept the terms linear in the sources because these terms may be used as a non-trivial check that we construct the correct path integral.

If we linearise in the sources then only the contribution of the first line of (\ref{QFT:1pt}) survives. This corresponds in gauge/gravity duality to linearising the bulk field equations. In this case the 1-point function is related to the 2-point function at the conformal point. Since 2-point functions in CFT are diagonal then the only operator that has a non-zero 1-point function is precisely the operator associated with the excited state
\begin{equation}
\langle \mathcal{O}_\Delta  \rangle \neq 0, \qquad \langle \mathcal{O}_i  \rangle =0 \qquad  {( linear\ approximation}).
\end{equation}
This implies that if we want to work out the linearised bulk solution dual to the state $|\Delta \rangle$, it suffices to only consider
the bulk field that is dual to the operator $\mathcal{O}_\Delta $ in a fixed AdS background.\footnote{Note that if we set $\phi_{(0)}^+=\phi_{(0)}^- \equiv \phi_{(0)}$ (with $\phi_{(0)}$ infinitesimal)  and the bulk action is quadratic in $\Phi$ so that the linear approximation is exact, the bulk solution would have the interpretation as being dual to the state $|0\rangle + \phi_{(0)} |\Delta\rangle$. In this paper we are taking the view that the bulk action contains interaction terms  and the linear approximation is the first step towards constructing the full solution perturbatively. From the full solution one may extract the in-in correlators in the state $|\Delta\rangle$ by computing the renormalised on-shell action and keeping the terms proportional to $\phi_{(0)}^+  \phi_{(0)}^-$.}

This is no longer the case if we consider the full field equations, as now the second line in (\ref{QFT:1pt}) is also relevant and
\begin{equation}
\langle \mathcal{O}_\Delta  \rangle \neq 0, \qquad \langle \mathcal{O}_i  \rangle \neq 0, 
\end{equation}
for all operators $\mathcal{O}_i$ that appear in the OPE of $\mathcal{O}_\Delta $ with
itself (so that the 3-point function in (\ref{QFT:1pt}) is non-zero). 
This implies that the bulk solution will now include all bulk fields that are dual to these operators.
In particular, the energy momentum tensor $T_{ij}$ appears in the OPE so one can no longer ignore the back-reaction to the metric.

The CFTs that appear in gauge/gravity duality admit a 't Hooft large $N$ limit and one may also use the large $N$ limit to organise the bulk reconstruction. In particular, if we normalise the operators such that their 2-point function is independent of $N$, then 3- and higher-point functions go to zero as $N \to \infty$. With this normalisation, the first line in (\ref{QFT:1pt}) is the leading order term in the large $N$ limit. We would like to emphasise however that with this normalisation not all $1/N^2$ terms correspond to non-planar corrections (quantum corrections in the bulk). 

An alternative normalisation is to normalise the operators such that all connected $n$-point functions scale as $N^2$ to leading order (i.e.~computed using planar diagrams). With this normalisation all $1/N^2$ corrections are associated with non-planar diagrams. In AdS/CFT this normalisation is known as the ``supergravity normalisation'': all leading order factors of $N$ come from Newton's constant and $1/N^2$ corrections are due to quantum corrections (loop diagrams).  
 
Either way the leading order construction of the bulk solution dual to a state is universal while the higher order terms depend on the CFT under consideration. In this paper we will discuss in detail the universal part of the construction. The method can be readily extended to higher order once the CFT input is given. 

To keep the technicalities at the minimum we will discuss the case of $2d$ CFT 
either on $R \times S^1$ (with coordinates $(t, \phi$)) or on $R^{1,1}$ (with coordinates $(t,x)$) and we set the source equal to one, $\phi_{(0)}^\pm =1$.   
For  a CFT on $R \times S^1$ the 1-point function in the first line in (\ref{QFT:1pt})  then gives,
\begin{eqnarray}
\langle \mathcal{O}_\Delta(t,\phi)  \rangle = \frac{C}{(\cos(t-i \epsilon) - \cos \phi)^{\Delta}}+\frac{C}{(\cos(t+i \epsilon) - \cos \phi)^{\Delta}},  \label{RS1}
\end{eqnarray}
 while for a CFT on $R^{1,1}$ we obtain
\begin{eqnarray}
\langle \mathcal{O}_\Delta(t,\phi)  \rangle = \frac{\tilde{C}}{(-(t-i \epsilon)^2 + x^2)^{\Delta}}+ \frac{\tilde{C}}{(-(t+i \epsilon)^2 +x^2)^{\Delta}}, \label{R11}
\end{eqnarray}
where $C$ and $\tilde{C}$ are the normalisations of the 2-point functions in the two cases\footnote{\label{normalization} Actually, since $R \times S^1$ and 
$R^{1,1}$ are conformally related one may relate (\ref{RS1}) and (\ref{R11}) and then $\tilde{C} = 2^\Delta C $ \cite{Skenderis2009}.}.
The bulk solution dual to this state in global AdS should reproduce (\ref{RS1}) while the  bulk solution in Poincar\'{e} AdS should yield (\ref{R11}).

\section{Global AdS} \label{sec:global}

As discussed in the previous section if we want to obtain the bulk solution dual to the state $|\Delta\rangle=\mathcal{O}_{\Delta} |0\rangle$  of a CFT on $R \times S^1$ to linear order in the sources, it suffices to consider a free scalar $\Phi$ of mass $m^2 =\Delta (\Delta-2)$ in global AdS -- this field is dual to the operator $\mathcal{O}_{\Delta}$. We will take $\Delta = 1 +l$ with $l=0,1,2,\ldots$,
as this is the case in most models embedded in string theory, though the results hold for any $\Delta \geq 1$ with minimal changes. We will also set $1/16 \pi G_N=1, \ell=1$, where $G_N$ is the three dimensional Newton constant and $\ell$ is the AdS radius.

The appropriate spacetime is that in the right panel of Fig. \ref{time contour and manifold}, with the Lorentzian pieces being global Lorenzian AdS spacetimes and the Euclidean ones, their Wick rotated version. The real-time gauge/gravity prescription instructs us to 
solve the field equations of the scalar  $\Phi$ in the four different parts of the spacetime and then match them. Since we are only aiming at constructing the leading order universal part, it suffices to solve the free field equations.

\subsection{Lorentzian Solution}\label{sec:GlobalSol}

The metric for global AdS$_{2+1}$ and for Lorentzian signature can be written as
\begin{equation}
\mathrm{d}s^2=-(1+r^2)\mathrm{d}t^2+\frac{\mathrm{d}r^2}{1+r^2}+r^2\mathrm{d}\phi^2. \label{globLorMetric}
\end{equation}
In these coordinates the conformal boundary of AdS is at $r \to \infty$.
The field equation describing a massive scalar field propagating in this background without back--reaction is given by
\begin{equation}
\lc (1+r^2)\partial_r^2 + \frac{1+3r^2}{r}\partial_r-\frac{1}{1+r^2}\partial_t^2+\frac{1}{r^2}\partial_\phi^2 -m^2 \rc \Phi(t, r,\phi) =0. \label{KG}
\end{equation}
Substituting the solution ansatz
\begin{equation} 
\mathrm{e}^{-i \omega t+i k \phi}f(\omega,k,r)
\end{equation}
one finds that $f(\omega, k, r)$ satisfies
\begin{equation}
0=(1+r^2)f''+ \frac{3r^2+1}{r} f'
- \left( \frac{k^2}{r^2} -\frac{\omega^2}{r^2+1}+m^2\right) f.  \label{radODE}
\end{equation}
where the prime denotes a derivative w.r.t. $r$.
The solution of this ODE is given in terms of a hypergeometric function,
\begin{equation}
f(\omega,k, r) = C_{\omega kl} (1+ r^2) ^{\omega/2} r^{|k|} \,_2F_1 (\hat{\omega}_{kl}, \hat{\omega}_{kl} -l; |k|+1;-r^2) \label{radmod}
\end{equation}
where $l=\Delta-1=\{0,1,2,\dots\}$, $\Delta = 1+\sqrt{1+m^2}$, $\hat{\omega}_{kl} = (\omega +|k| + l +1)/2, \,  k \in \mathbb{Z}, \, \omega \in \mathbb{R}$ and $C_{\omega k l} = (\Gamma( \hat{\omega}_{kl}) \Gamma(\hat{\omega}_{kl} - \omega))/((l-1)! \, |k|!)$. 
The normalisation constant has been chosen to make the coefficient of the leading order term in the near boundary expansion of $f(\omega,k,r)$ equal to 1. Note that $f(\omega, k,r)=f(\omega,-k,r) = f(\omega,|k|r)$ and $f(\omega, k,r) = f(-\omega, k, r)$.

Near the conformal boundary the solution admits the following series expansion in $r$,
\begin{equation} 
f(\omega, |k|, r) = r^{l-1} + \dots + r^{-l-1} \alpha(\omega, |k|, l) \left[ \ln(r^2) + \beta( \omega, |k|, l) \right] + \dots \label{asymp}
\end{equation}
where 
\begin{subequations}
\begin{align}
\alpha(\omega, |k|, l ) &= \frac{ (\hat{\omega}_{kl} - l)_l ( \hat{\omega}_{kl} -|k| - l)_l}{l! \, (l-1)!} \\[5pt]
\beta( \omega, |k|, l) &= - \psi( \hat{\omega}_{kl}) - \psi ( \hat{\omega}_{kl} - l - \omega).
\end{align}
\end{subequations}
From this expression we see that the modes have simple poles in the $\omega$ plane which appear at normalisable order, i.e.~at $r^{-l-1} = r^{-\Delta}$.
Thus, by integrating over $\omega$, in the absence of sources, we obtain the normalisable modes. 

The poles of $f(\omega,k,r)$ are at  $\omega = \omega_{nk}^\pm = \pm (2n+|k|+l+1), \, n \in \mathbb{N}$.
It follows that near the conformal boundary the normalisable modes are given by
\begin{align}
g(\omega_{nk}, |k|, r) =&\frac{1}{4\pi^2 i} \oint_{\omega_{nk}} \!\!\! d\omega \, f(\omega, |k|, r) \nonumber \\
=& \frac{1}{4\pi^2 i } \oint_{\omega_{nk}}\! \!\! d\omega \bigg[ \text{non-norm. term} + \frac{(\hat{\omega}_{kl}-l)_l \, (\hat{\omega}_{kl}-|k|-l)_l}{l!(l-1)!}\big(\ln(r^2)\nonumber \\
&\phantom{\oint_{\omega_{nk}}\! \!\! d\omega \Big[ } - \psi(\hat{\omega}_{kl}) - \psi(\hat{\omega}_{kl}-\omega-l) \big) + \dots \bigg] \nonumber \\
=& \frac{1}{\pi} r^{-l-1} \frac{(n+|k|+1)_l (n+l)!}{n!l!(l-1)!} + \dots  \label{asymptoticGlobal}
\end{align}
where the contours are defined clockwise for the poles at $\omega_{nk}^+$ and counterclockwise for poles at $\omega_{nk}^-$ such that $g(\omega_{nk}^+,|k|,r) = g(\omega_{nk}^-,|k|,r)$.
Combining this result with equation (\ref{radmod}) allows us to extend the normalisable modes to finite $r$,
\begin{align}{rl}
g(\omega_{nk}, |k|, r)= &\frac{1}{\pi} \, r^{|k|} (1+r^2)^{-\frac{|k|+l+1}{2}} \frac{(n+1)_l (n+|k|+1)_l}{l! (l-1)!} \nn \\
& \,_2F_1\left(n+|k|+l+1,-n;l+1;\frac{1}{1+r^2}\right).\;  \label{fullGlobal}
\end{align}
Then, a normalisable Lorentzian solution has the form
\begin{equation} \label{eq:normalizable}
\Phi_L(t,r,\phi)=\sum_{k \in \mathbb{Z}} \sum_{n=0}^\infty \left( b_{nk} \,\mathrm{e}^{-i \omega_{nk}^+ t+ik\phi}+b_{nk}^\dagger \, \mathrm{e}^{-i \omega_{nk}^- t-ik \phi}\right)g(\omega_{nk},|k|,r), \quad
\end{equation}
where $b_{nk}$ and $b_{nk}^\dagger$ are arbitrary coefficients, to be determined from the matching conditions.

\subsection{Euclidean Solution}

The metric for global AdS$_{2+1}$ and for Euclidean signature can be obtained from the Lorentzian one, (\ref{globLorMetric}), by Wick rotation, $t=-i\tau$.
Similarly, one may obtain the Euclidean solutions by analytically continuing the Lorentzian modes,
\begin{align}
\mathrm{e}^{-\omega \tau+i k \phi} f(\omega,k, r) =& C_{\omega kl} \, \mathrm{e}^{-\omega \tau +i k \phi} (1+ r^2) ^{\omega/2} r^{|k|} \nonumber \\
&\,_2F_1 (\hat{\omega}_{kl}, \hat{\omega}_{kl} -l; |k|+1;-r^2).
\end{align}
In accordance with our choice of boundary conditions, the general solution in the Euclidean caps requires that we turn on a source $\phi_{(0)}(\tau, \phi)$ on the boundary. 
Since we are working with momentum modes, we need to express the source in momentum space.
For a general source $\phi_{(0)}^-(\tau, \phi)$ with support on the boundary of the past Euclidean cap and away from the matching surface at $\tau = 0$ we have
\begin{equation} \label{source}
\phi_{(0)}^-(\omega, k) = \int_0^{2\pi} \!\!\! \d \phi \int^{0}_{-\infty} \!\!\! \d \tau \, \mathrm{e}^{\omega \tau - i k \phi} \phi_{(0)}^-(\tau, \phi)
\end{equation}
Since the range of $\tau$ is over the half real line only, it is natural to use Laplace rather than Fourier transforms.
Using this, the most general solution in the past Euclidean cap is
\begin{align}
\Phi_E^-(\tau, r, \phi) 
=&\frac{1}{4\pi^2 i} \sum_{k \in \mathbb{Z}}\int^{i\infty}_{-i\infty}\!\!\! d\omega \, \mathrm{e}^{-\omega \tau + i k \phi} \phi_{(0)}^-(\omega, k) f(\omega, |k|, r) \nn \\
& + \sum_{k \in \mathbb{Z}} \sum_{n=0}^\infty  d_{nk}^- \mathrm{e}^{-\omega_{nk}^- \tau + i k \phi} g(\omega_{nk}, k, r)\label{past_sol} 
\end{align}
where the integration over $\omega$ is along the imaginary axis
and $g(\omega_{nk},|k|,r)$ is defined in (\ref{fullGlobal}). 
The second term in equation (\ref{past_sol}) is included to make the solution as general as possible. It behaves as $r^{-l-1}$ near the boundary and it decays exponentially as $\tau \to -\infty$ so it does not affect the asymptotic behaviour of the solution and, therefore, it can not be excluded. 

To explicitly see that the solution has a source term, recall that for large $r$, $f$ has the expansion in (\ref{asymp}) and thus the Euclidean solution asymptotes to\footnote{Here we assume that the source admits a Laplace transform.  This is true in particular if 
$\phi_{(0)}(\omega,k)$ can be extended to a meromorphic function with no singularities for Re$(\omega) > c$, for some finite $c$.
Here for simplicity we take $c=0$.}
\begin{align}
\Phi_E^-(\tau, r, \phi) =& r^{l-1} \frac{1}{4\pi^2 i} \sum_{k \in \mathbb{Z}}\int^{i\infty}_{-i\infty}\!\!\! d\omega \, \mathrm{e}^{-\omega \tau + i k \phi} \phi_{(0)}^-(\omega, k) + O(r^{l-2}) \nonumber \\
 =& r^{l-1} \phi_{(0)}(\tau,\phi) +  O(r^{l-2}) 
\end{align}
In this paper we choose the source profile to be a $\delta$--function localised at $(\tau, \phi) = (-\epsilon,0)$, $\epsilon>0$, i.e.~$\phi_{(0)}^-\left(\tau, \phi\right) = \delta(\tau+\epsilon) \delta(x)$, which implies $\phi_{(0)}^-(\omega,k)= \exp (-\omega \epsilon)$.

The integral over $\omega$ can be done explicitly close to the matching surface using contour integration.
Denoting time in the past Euclidean cap by $\tau_0$ and considering $-\epsilon <\tau_0\leq 0$ we close 
 the $\omega$--contour to the right (such that $\text{Re}(\omega)>0$), and picking up the contributions from the poles at $\omega=\omega_{nk}^+$ we obtain
\begin{align}\label{past_sol_fin}
\Phi_E^- (\tau_0, r, \phi) =\sum_{k \in \mathbb{Z}} 
\sum_{n=0}^\infty &\left(
\phi_{(0)}^-(\omega_{nk}^+, k) \mathrm{e}^{-\omega_{nk}^+ \tau_0 + i k \phi} 
\right.\nonumber \\ &\left. 
+d_{nk}^- \mathrm{e}^{-\omega_{nk}^- \tau_0 + i k \phi} \right)
g(\omega_{nk}, |k|, r). 
\end{align}

The analysis for the future Euclidean cap follows along the same lines. 
In particular, denoting Euclidean time in the future Euclidean cap by $\tau_3$, $0\leq \tau_3 < \infty$,  and 
using a $\delta$--function source localised at $(\tau_3, \phi) =(\epsilon,0)$ where $\epsilon$ is the same as for the past Euclidean cap, $\phi_{(0)}^+(\tau_3,\phi) = \delta(\tau_3 -\epsilon)\delta(\phi), \ \phi_{(0)}^+(\omega,k)= \exp (\omega \epsilon)$ and considering the solution close to the matching surface, $0 \leq \tau_3 < \epsilon$, 
we obtain
\begin{align} \label{fut_sol_fin}
\Phi_E^+ (\tau_3, r, \phi) = \sum_{k \in \mathbb{Z}} \sum_{n=0}^\infty  & \left( 
 \phi_{(0)}^+(\omega_{nk}^-, k) \mathrm{e}^{-\omega_{nk}^- \tau_3 + i k \phi} 
\right.\nonumber \\ &\left. 
+ \tilde{d}_{nk}^+ \mathrm{e}^{-\omega_{nk}^+ \tau_3 + i k \phi} \right) g(\omega_{nk}, |k|, r). \quad
\end{align}

\subsection{Matching Conditions} \label{contour_discussion}

 The time contour considered here is the in--in contour shown on the left of figure \ref{fig:inincontour}, with the corresponding AdS manifold shown on the right.
It runs from $i \infty$ to $0$, then to $T$, then back to $0$ and then to $-i\infty$. Accordingly, the contour--integrated action is
\begin{align}
S =& -\int_{-\infty}^0 \!\mathrm{d} \tau_0 \; L_E(\Phi_E^-) + i \int_0^T \!\! \mathrm{d} t_1 \; L_L(\Phi_L^1) - i \int_T^{2T} \!\! \mathrm{d}t_2 \; L_L (\Phi_L^2 )\nonumber \\
& - \int_0^\infty \!\! \mathrm{d} \tau_3 \; L_E(\Phi_E^+ ).
\end{align}
where 
\begin{equation}
L_E =\tfrac{1}{2} \int \! \mathrm{d}^3 x \sqrt{g} \left(g^{\mu \nu} \partial_\mu \Phi_E \partial_\nu \Phi_E + m^2 \Phi_E^2\right)
\end{equation}
 and 
 \begin{equation}
 L_L = -\tfrac{1}{2} \int\mathrm{d}^3 x \sqrt{-g} \left(g^{\mu \nu} \partial_\mu \Phi_L\partial_\nu \Phi_L - m^2 \Phi_L^2\right).
 \end{equation}

\begin{center}
\begin{figure}[t]
\centering
\begin{tikzpicture}[line width = 0.2mm, scale=0.6]
\draw (0,3) -- (0,0.1) -- (6, 0.1) -- (6,-0.1) --(0,-0.1) -- (0, -3);
\node [anchor= west] at (0.2, 2) {$\tau_0$};
\node [anchor = west] at (0.2cm, -2 cm) {$\tau_3$};
\node [anchor = south] at (3 cm, 0.35 cm) {$t_1$};
\node [anchor = north] at (3 cm, -0.35 cm) {$t_2$};
\node [anchor =south west] at (6 cm,0 cm) {$T$};
\draw[->] (0.25, 2.5) -- (0.25,1.5);
\draw[->] (0.25, -1.5) -- (0.25, -2.5);
\draw[line width =0.2 mm, ->] (2.5 cm,0.35 cm) -- (3.5 cm, 0.35 cm); 
\draw[line width =0.2 mm, ->] (3.5 cm,-0.35 cm) -- (2.5 cm, -0.35 cm); 
\node [anchor=center] at ( 0cm,1.1 cm) {\small$\times$};
\node [anchor=center] at (0cm, -1.1 cm) {\small $\times$};
\node [anchor = east] at (0 cm, 1.1 cm) {$\epsilon$};
\node [anchor = east] at (0 cm, -1.1 cm) {$-\epsilon$};
\draw [line width = 0.3mm, ->] (8.8, 0) -- (11.2,0);
\draw[line width=0.2 mm ,dashed] (14 cm ,3 cm) arc (-180:-360:2 cm and 0.7 cm);
\draw[line width=0.2 mm] (14 cm ,3 cm) arc (180:360:2 cm and 0.7 cm);
\draw[line width = 0.2mm] (14 cm, 3 cm ) -- (14 cm, -3 cm);
\draw[line width=0.2 mm ,dashed] (14 cm ,0 cm) arc (-180:-360:2 cm and 0.7 cm);
\draw[line width=0.2 mm] (14 cm ,0 cm) arc (180:360:2 cm and 0.7 cm);
\draw[line width = 0.2mm] (18 cm, 3 cm ) -- (18 cm, -3cm);
\draw[line width=0.2 mm ,dashed] (14 cm , -3 cm) arc (-180:-360:2 cm and 0.7 cm);
\draw[line width=0.2 mm] (14 cm , -3 cm) arc (180:360:2 cm and 0.7 cm);
\draw[line width=0.2mm] (14 cm, 3 cm) arc (180:0:2 cm);
\draw[line width=0.2mm, ->] (13.7 cm, 3 cm) arc (180:145:2.3 cm);
\node at (13.6 cm, 4 cm) {$\tau_3$};
\draw[line width=0.2mm, ->] (13.7 cm, 1 cm) -- (13.7, 2 cm);
\node at (13.3 cm, -1.5 cm) {$t_1$};
\node at (16 cm, 1.5 cm) {$L$};
\draw[line width = 0.2mm, ->] (13.7 cm, -2 cm) -- (13.7 cm, -1 cm);
\node at (13.3 cm, 1.5 cm) {$t_2$}; 
\node [anchor = center] at (16 cm, -1.5 cm) {$L$};
\draw[line width=0.2mm] (14 cm, -3 cm) arc (180:360:2 cm);
\node at (18.35 cm, 0 cm) {$T$};
\node at (18.5 cm, 3 cm) {$2T$};
\draw[line width=0.2mm, <-] (13.7 cm, -3 cm) arc (180:225:2.3 cm);
\node at (13.6 cm, -4 cm) {$\tau_0$};
\node [anchor= center] at (16 cm, 4.2 cm) {$E$};
\node at (16 cm, -4.2 cm) {$E$};
\node [anchor =  south west] at (17.7 cm, 4 cm) {$\epsilon$};
\node [anchor = north west] at (17.75 cm, -4 cm) {$-\epsilon$};
\node at (17.7 cm, 4 cm) {$\times$};
\node at (17.75 cm, -4 cm) {$\times$};
\end{tikzpicture}
\caption{In--in time contour (left) and corresponding AdS manifold (right).}
\label{fig:inincontour}
\end{figure}
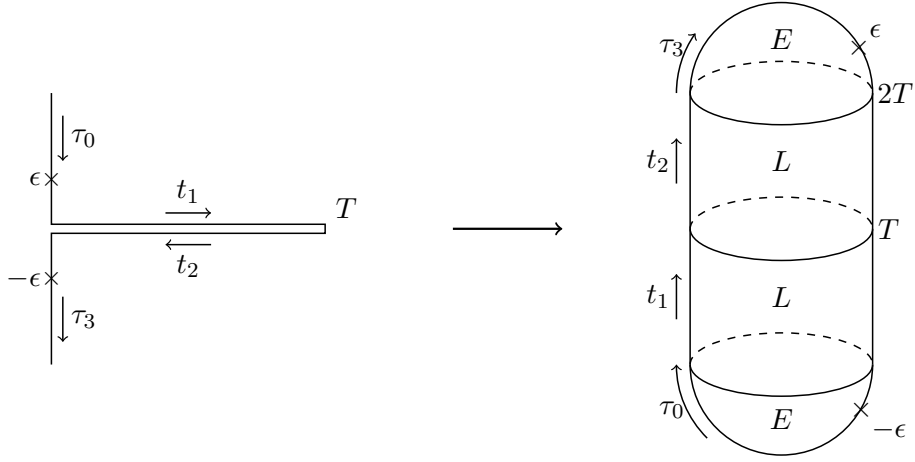
\end{center}

The matching conditions are
\begin{align}
\Phi_E^-\big|_{\tau_0 = 0} &= \Phi_L^1 \big|_{t_1 = 0}, \qquad \qquad & \partial_{\tau_0} \Phi_E^- \big|_{\tau_0=0} &= -i \partial_{t_1} \Phi_L^1 \big|_{t_1 = 0} \nonumber \\
\Phi_L^1 \big|_{t_1 = T} &= \Phi_L^2 \big|_{t_2 = T}, &  \partial_{t_1} \Phi_L^1 \big|_{t_1 = T} &= - \partial_{t_2} \Phi_L^2 \big|_{t_2 = T}  \label{matchingconditions} \\
\Phi_L^2 \big|_{t_2 = 2T} &= \Phi_E^+ \big|_{\tau_3 =0}, & \partial_{t_2} \Phi_L^2 \big|_{t_2 =2T} &= -i\partial_{\tau_3} \Phi_E^+ \big|_{\tau_3=0}. \nonumber
\end{align}

From the previous section we have that the solutions in the four manifolds are
\begin{subequations}
\begin{flalign}
&-\epsilon<\tau_0 \leq 0:&  \nonumber \\
&\hspace{15pt}\Phi_E^- (\tau_0, r, \phi) =\sum_{k \in \mathbb{Z}} 
\sum_{n=0}^\infty  \left(
\phi_{(0)}^-(\omega_{nk}^+, k) \mathrm{e}^{-\omega_{nk}^+ \tau_0 + i k \phi} 
+d_{nk}^- \mathrm{e}^{-\omega_{nk}^- \tau_0 + i k \phi} \right)
 g(\omega_{nk}, |k|, r) &\nonumber \\
&&\\
&0 \leq \tau_3 < \epsilon: &\nonumber \\ 
&\hspace{15pt} \Phi_E^+ (\tau_3, r, \phi) = \sum_{k \in \mathbb{Z}} \sum_{n=0}^\infty   \left( 
 \phi_{(0)}^+(\omega_{nk}^-, k) \mathrm{e}^{-\omega_{nk}^- \tau_3 + i k \phi} 
+\tilde{d}_{nk}^+ \mathrm{e}^{-\omega_{nk}^+ \tau_3 + i k \phi} \right) g(\omega_{nk}, |k|, r)& \nonumber \\ &&\\
&0 \leq t_1 \leq T :& \nonumber  \\
&\hspace{15pt} \Phi_L^1(t_1,r,\phi)=\sum_{k \in \mathbb{Z}} \sum_{n=0}^\infty \left( b_{nk} \,\mathrm{e}^{-i \omega_{nk}^+ t_1+ik\phi}+b_{nk}^\dagger \, \mathrm{e}^{-i \omega_{nk}^- t_1-ik \phi}\right)g(\omega_{nk},|k|,r), \nonumber &\\
\label{norm1a}& \\
&T\leq t_2 \leq 2T : \nonumber \\
&\hspace{15pt}  \Phi_L^2(t_2,r,\phi)=\sum_{k \in \mathbb{Z}} \sum_{n=0}^\infty \left( \tilde{b}_{nk} \,\mathrm{e}^{-i \omega_{nk}^+ t_2+ik\phi}+\tilde{b}_{nk}^\dagger \, \mathrm{e}^{-i \omega_{nk}^- t_2-ik \phi}\right)g(\omega_{nk},|k|,r).& \nonumber \\
\end{flalign}
\end{subequations}

 Applying to these the matching conditions 
we obtain the following relations:
from the matching conditions at $\tau_0 = 0, t_1 = 0$ 
\begin{subequations}
\begin{align}
b_{nk} =& 
 \phi_{(0)}^-(\omega_{nk}^+,k) =
 \mathrm{e}^{-\omega_{nk}^+\epsilon}, \label{sub1a} \\
b_{nk}^\dagger =& d_{nk}^- . \label{sub2}
\end{align}
\end{subequations}

From the matching conditions at $t_1=T, t_2=T$ 
\begin{subequations}
\begin{align}
b_{nk}^\dagger =& \tilde{b}_{nk}\mathrm{e}^{-2i\omega_{nk}^+ T} ,\label{sub3} \\
b_{nk} =& \tilde{b}^\dagger_{nk}\mathrm{e}^{-2i \omega_{nk}^- T}. \label{sub4}
\end{align}
\end{subequations}
Finally, from the matching conditions at $t_2 =2T, \tau_3 = 0$ 
\begin{subequations}
\begin{align}
 \tilde{b}_{nk}=&
 \phi_{(0)}^+(\omega_{nk}^-,k) \mathrm{e}^{-2i\omega_{nk}^-T} =
 \mathrm{e}^{-i\omega_{nk}^-(2T+i \epsilon)}, \label{sub5} \\
\tilde{b}_{nk}^\dagger =&\tilde{d}^+_{nk} \mathrm{e}^{-2i \omega_{nk}^+T}. \label{sub6}
\end{align}
\end{subequations}
Note that had we chosen the position in complex time where we insert the sources to be different for the two caps, say $\tau_{0,\text{source}}=-\epsilon$ and $\tau_{3,\text{source}}=\tilde{\epsilon}$, where $\tilde{\epsilon}>0$, then the relationships $b_{nk} =\left( b_{nk}^{\dagger}\right)^*$ and $\tilde{b}_{nk} = \left(\tilde{b}_{nk}^{\dagger}\right)^*$would have implied that $\epsilon = \tilde{\epsilon}$.

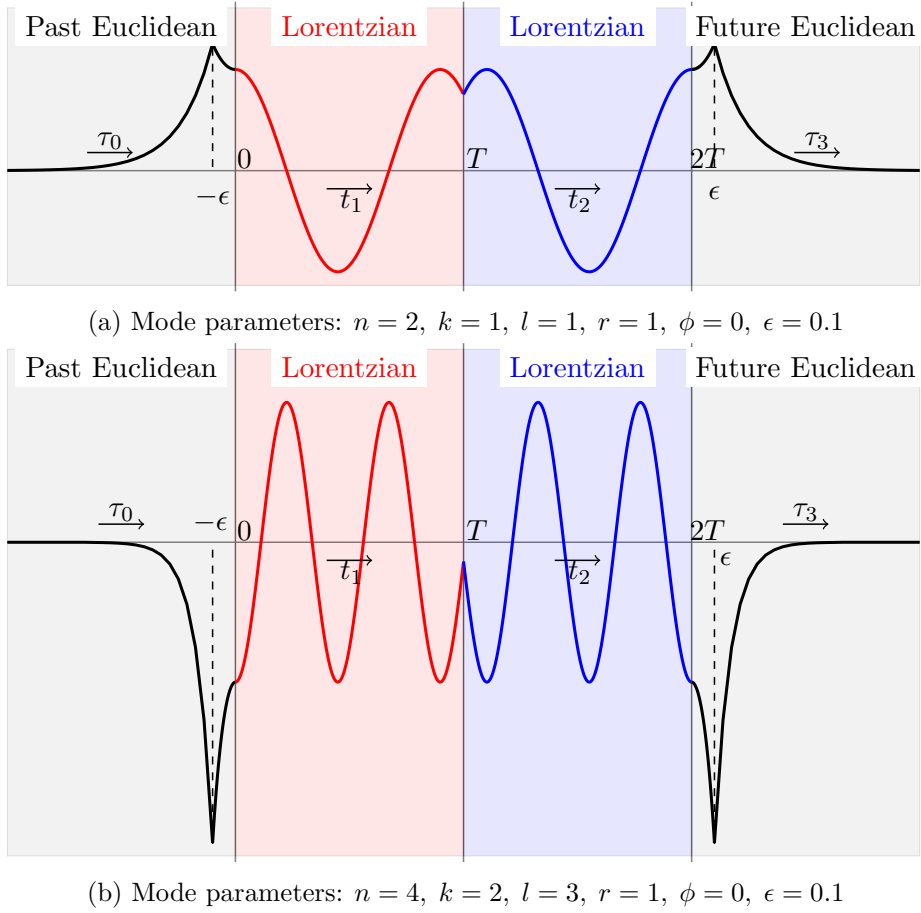
\begin{figure}[t!]
\centering
\begin{subfigure}{\textwidth}
\centering
\begin{tikzpicture}[xscale =3, yscale=8]
\draw [help lines, -, line width = 0.2mm] (-1,0) -- (3,0);
\draw [help lines, line width = 0.2mm] (0,-0.2) -- (0,0.28);
\draw [help lines, line width = 0.2mm] (1,-0.2) -- (1,0.28);
\draw [help lines, line width = 0.2mm] (2,-0.2) -- (2,0.28);
\draw [fill=gray, opacity=0.1] (-1, -0.19) rectangle (0, 0.27);
\draw [fill=red, opacity=0.1] (0, -0.19) rectangle (1, 0.27);
\draw [fill=blue, opacity=0.1] (1, -0.19) rectangle (2, 0.27);
\draw [fill=gray, opacity=0.1] (2, -0.19) rectangle (3, 0.27);
\draw [dashed, line width = 0.2mm] (-0.1,0.23) -- (-0.1,0);
\draw [dashed, line width = 0.2mm] (2.1,0.23) -- (2.1,0);
\node[below] at (2.1,-0.01) {$\epsilon$};
\node[below] at (-0.1,-0.01) {$-\epsilon$};
\draw [black, line width=0.4mm, domain=-1:-0.1] plot (\x, {0.0838282 * exp(7* \x) +0.33994 * exp(7 *\x)});
\draw [black, line width = 0.4mm, domain=-0.1:0] plot (\x, {0.0838282 * exp(7 * \x ) + 0.0838282 * exp(-7* \x )}); 
\draw [line width =0.2mm, ->] (-0.65,0.03)--(-0.45,0.03);
\node at (-0.55, 0.05) {$\tau_0$};
\node at (0.04, 0.02) {$0$};
\node[rectangle, fill=white] at (-0.5,0.24) {Past Euclidean};
\draw [red, line width=0.4mm, samples = 700, domain=0:1] plot(\x, {0.167656 * cos(7*\x r)});
\draw [line width =0.2mm, ->] (0.4,-0.03)--(0.6,-0.03);
\node at (0.51, -0.05) {$t_1$};
\node at (1.06, 0.02) {$T$};
\node[rectangle, fill=white] at (0.5,0.24) {\textcolor{red}{Lorentzian}};
\draw [blue, line width=0.4mm,samples = 700, domain=1:2] plot(\x,{0.167656 * cos(14 r- 7*\x r)});
\draw [line width =0.2mm, ->] (1.4,-0.03)--(1.6,-0.03);
\node at (1.51, -0.05) {$t_2$};
\node at (2.07, 0.02) {$2T$};
\node[rectangle, fill=white] at (1.5,0.24) {\textcolor{blue}{Lorentzian}};
\draw [black, line width=0.4mm,samples = 700, domain =0:0.1] plot(\x+2,{0.0838282* exp(-7* \x) +0.0838282  *exp(7*\x)});
\draw [black, line width = 0.4mm, domain=0.1:1] plot (\x+2,{0.0838282* exp(-7* \x)+0.33994*exp(-7*\x)});
\draw [line width =0.2mm, ->] (2.45,0.03)--(2.65,0.03);
\node at (2.55, 0.05) {$\tau_3$};
\node[rectangle, fill=white] at (2.5,0.24) {Future Euclidean};
\end{tikzpicture}
\caption{Mode parameters: $n=2,\; k=1,\; l=1,\; r=1,\;  \phi = 0,\; \epsilon = 0.1$}
\end{subfigure}
\vspace{25pt}
\centering
\begin{subfigure}{\textwidth}
\centering
\begin{tikzpicture}[xscale=3, yscale = 0.8]
\draw [help lines, -, line width = 0.2mm] (-1,0) -- (3,0);
\draw [help lines, line width = 0.2mm] (0,-5.3) -- (0,3.3);
\draw [help lines, line width = 0.2mm] (1,-5.3) -- (1,3.3);
\draw [help lines, line width = 0.2mm] (2,-5.3) -- (2,3.3);
\draw [fill=gray, opacity=0.1] (-1, -5.2) rectangle (0, 3.2);
\draw [fill=red, opacity=0.1] (0, -5.2) rectangle (1, 3.2);
\draw [fill=blue, opacity=0.1] (1, -5.2) rectangle (2, 3.2);
\draw [fill=gray, opacity=0.1] (2, -5.2) rectangle (3, 3.2);
\draw [dashed, line width = 0.2mm] (-0.1,-5) -- (-0.1,0);
\draw [dashed, line width = 0.2mm] (2.1,-5) -- (2.1,0);
\node[below] at (2.15,0) {$\epsilon$};
\node[above] at (-0.11,0.0) {$-\epsilon$};
\draw [black, line width=0.4mm, domain=-1:-0.1] plot (\x, {-1.15902 * exp(14* \x) -19.0596 * exp(14 *\x)});
\draw [black, line width = 0.4mm, domain=-0.1:0] plot (\x, {-1.15902 * exp(14 * \x ) - 1.15902 * exp(-14* \x )}); 
\draw [line width =0.2mm, ->] (-0.6,0.3)--(-0.4,0.3);
\node at (-0.5, 0.5) {$\tau_0$};
\node at (0.04, 0.2) {$0$};
\node[rectangle, fill=white] at (-0.5,2.9) {Past Euclidean};
\draw [red, line width=0.4mm, samples = 700, domain=0:1] plot(\x, {-2.31803 * cos(14*\x r)});
\draw [line width =0.2mm, ->] (0.4,-0.3)--(0.6,-0.3);
\node at (0.51, -0.5) {$t_1$};
\node at (1.06, 0.2) {$T$};
\node[rectangle, fill=white] at (0.5,2.9) {\textcolor{red}{Lorentzian}};
\draw [blue, line width=0.4mm,samples = 700, domain=1:2] plot(\x,{-2.31803 * cos(28 r- 14*\x r)});
\draw [line width =0.2mm, ->] (1.4,-0.3)--(1.6,-0.3);
\node at (1.51, -0.5) {$t_2$};
\node at (2.07, 0.2) {$2T$};
\node[rectangle, fill=white] at (1.5,2.9) {\textcolor{blue}{Lorentzian}};
\draw [black, line width=0.4mm,samples = 700, domain =0:0.1] plot(\x+2,{-1.15902* exp(-14* \x) -1.15902  *exp(14*\x)});
\draw [black, line width = 0.4mm, domain=0.1:1] plot (\x+2,{-1.15902* exp(-14* \x)-19.0596 *exp(-14*\x)});
\draw [line width =0.2mm, ->] (2.4,0.3)--(2.6,0.3);
\node at (2.5, 0.5) {$\tau_3$};
\node[rectangle, fill=white] at (2.5,2.9) {Future Euclidean};
\end{tikzpicture}
\caption{Mode parameters: $n=4,\; k=2,\; l=3,\; r=1,\;  \phi = 0,\; \epsilon = 0.1$}
\end{subfigure}
\caption{Tracing individual modes through the four segments of the manifold.}
\label{plot of modes}
\end{figure}

 In what follows we refer to terms proportional to $\mathrm{e}^{-i\omega_{nk}^+ t}$ ($\mathrm{e}^{\omega_{nk}^+\tau}$ for Euclidean) as the positive frequency modes and $\mathrm{e}^{-i\omega_{nk}^- t}$ ($\mathrm{e}^{-\omega_{nk}^- \tau}$ for Euclidean) as the negative frequency modes.
From the matching conditions we observe that the positive frequency exponential source modes from the past Euclidean cap source the positive frequency oscillatory normalisable modes in the first Lorentzian manifold. As these modes evolve into the second Lorentzian manifold they give rise to the negative frequency oscillatory normalisable modes. Finally, they become positive frequency normalisable modes in the future Euclidean cap.
 The negative frequency source modes from the past Euclidean manifold decay and do not enter the Lorentzian manifolds.
In addition to source modes, there are negative frequency normalisable modes in the past Euclidean manifold. 
 These modes come from negative frequency source modes in the future Euclidean cap which become positive frequency normalisable modes in the second Lorentzian manifold, then evolve into negative frequency normalisable modes in the first Lorentzian manifold and finally they give rise to negative normalisable modes in the past Euclidean cap.
The absence of positive frequency normalisable modes in the past Euclidean manifold is due to the fact that these grow exponentially as $\tau_0 \to -\infty$.
 Schematically, the different modes evolved as shown below: 
Starting from the past Euclidean modes,
\begin{align}
&\phi_{0}^-(\omega_{nk}^+,k) \longrightarrow b_{nk} \longrightarrow \tilde{b}_{nk}^\dagger\mathrm{e}^{-2i\omega_{nk}^-T} \longrightarrow \tilde{d}_{nk}^+& \nn \\[5pt]
&\phi_{(0)}^-(\omega_{nk}^-,k) \longrightarrow \text{decay} &  \\[5pt]
&d_{nk}^- \longrightarrow  b_{nk}^\dagger \longrightarrow \tilde{b}_{nk} \mathrm{e}^{-2i \omega_{nk}^+T}  \longrightarrow 
\phi_{(0)}^+(\omega_{nk}^-,k),&\nn
\end{align}
and, similarly, starting from the future Euclidean cap,
\begin{align}
&\phi_{(0)}^+(\omega^-_{nk},k) \longrightarrow \tilde{b}_{nk} \mathrm{e}^{-2i \omega_{nk}^+ T} \longrightarrow b_{nk}^\dagger \longrightarrow d_{nk}^- \nn& \\[5pt]
&\phi_{(0)}^+(\omega_{nk}^+,k) \longrightarrow \text{decay}&   \\[5pt]
&\tilde{d}_{nk}^+  \longrightarrow  \tilde{b}_{nk}^\dagger \mathrm{e}^{-2i\omega_{nk}^-T}\longrightarrow b_{nk} \longrightarrow 
 \phi_{(0)}^-(\omega_{nk}^+,k).\nn&
\end{align}
Figure \ref{plot of modes} shows plots of the time evolution of individual modes from exponentially decaying source modes in the Euclidean manifolds to oscillatory, normalisable modes in the Lorentzian manifolds.
These plots were obtained by fixing $r$ and $\phi$ to be $1$ and $0$ respectively, and with the source insertions located at $\epsilon = 0.1$. 
The vertical axis corresponds to the amplitude of the scalar mode and the horizontal axis to contour time. 
Then these plots show two individual modes as they evolve from imaginary time in the past Euclidean manifold, to real time in the two Lorentzian manifolds and then back to imaginary time in the future Euclidean manifolds.

Combining all three sets of relationships between the coefficients of the different modes we find
\begin{subequations}
\begin{align}
b_{nk}^\dagger =&
\phi_{(0)}^+(\omega_{nk}^-,k),\\ 
\tilde{b}_{ nk}^\dagger =& 
 \phi_{(0)}^-(\omega_{nk}^+,k) \mathrm{e}^{-2i\omega_{nk}^+ T}.
\end{align}
\end{subequations}

Returning to the Lorentzian fields, we can now replace the original, arbitrary coefficients $b_{nk}^\pm$ and $\tilde{b}_{nk}^\pm$ with the above results to obtain expressions in terms of the Euclidean source modes.
\begin{subequations}
\begin{alignat}{3}
\Phi_L^1(t, r, \phi)\!=&\! 
\sum_{n=0}^\infty \sum_{k \in \mathbb{Z}}&&\left[\phi_{(0)}^-(\omega_{nk}^+,k) \mathrm{e}^{-i\omega_{nk}^+ t+ik\phi} \right.\nonumber \\
 &&&\left. \!+\phi_{(0)}^+(\omega_{nk}^-,k) \mathrm{e}^{-i\omega_{nk}^- t-ik\phi}\right]\! g(\omega_{nk}, |k|, r) \label{eq:20}\\[5pt]
 \Phi_L^2 (t, r, \phi)\!=&
 \sum_{n=0}^\infty  \sum_{k \in \mathbb{Z}}&& \left[ \phi_{(0)}^+(\omega_{nk}^-,k) \mathrm{e}^{i\omega_{nk}^+t+ik\phi}\right. \nonumber \\
 &&&\left.\! +\phi_{(0)}^-(\omega_{nk}^+,k) \mathrm{e}^{i\omega_{nk}^- t -ik\phi} \right]\! g(\omega_{nk}, |k|, r). \label{eq:23}
\end{alignat}
\end{subequations}
 where we used the relation between physical and contour time, $t_1=t$ and $2 T -t_2=t$.
  
\subsection{1-point function} \label{BoundaryinterpretationGlobal}

Having constructed normalisable Lorentzian solutions, we will now extract the 1-point function to verify that this  solution is indeed dual to the state $|\Delta \rangle$. For this we need to obtain the asymptotic expansion of the bulk field near the conformal infinity as in (\ref{as_exp}) and use~\cite{DeHaro2000},
\begin{equation}
\langle \mathcal{O}_\Delta(t,\phi) \rangle =- (2\Delta -2) \phi_{(2\Delta -d)} (t,\phi). \label{eq:field-1pt}
\end{equation}
We can choose to consider the insertion either in the upper part of the contour or in the lower. In the former case the 1-point function can be extracted from the asymptotic expansion of $\Phi_L^1$ while in the latter case from the asymptotic expansion of $\Phi_L^2$. In both cases, the answer should be the same. 

For concreteness, we consider the case the operator is in the upper part of the contour so the relevant field is $\Phi_L^1$.
Since this a normalisable mode, $\phi_{(2 \Delta -2)}$ is the coefficient of the leading order term as $r \to \infty$,
\begin{equation}
\phi_{(2\Delta -2)} =\frac{1}{\pi} \sum_{n=0}^\infty \sum_{k \in \mathbb{Z}} \mathrm{e}^{-\omega_{nk}^+ \epsilon}\left(\mathrm{e}^{-i\omega_{nk}^+t+ik\phi} +\mathrm{e}^{-i\omega_{nk}^- t -ik\phi}\right)
 \alpha(\omega_{nk},|k|,l),  
\end{equation}
where we have used
\begin{equation}
g(\omega_{nk}, |k|,r) = \frac{1}{\pi}r^{-\Delta} \alpha(\omega_{nk},|k|, l) +  \mathrm{O}\big(r^{-\Delta-1} \big)
\end{equation}
Performing the sums over $n$ and $k$ and inserting in (\ref{eq:field-1pt})
we finally get
\begin{equation}
\langle \mathcal{O}_\Delta(t,\phi)\rangle = \frac{l^2}{2^l \pi} \left(\frac{1}{(\cos(t-i \epsilon) - \cos \phi)^{\Delta}}+\frac{1}{(\cos(t+i \epsilon) - \cos \phi)^{\Delta}}\right). \label{1ptfunctionGlobal}
\end{equation}
This is indeed equal to value we got via a QFT computation in (\ref{RS1}). In our case, $C=l^2/(2^l \pi)$, which  is the standard supergravity normalisation of the 2-point function.
 
\section{Poincar\'e AdS} \label{sec:poincare}

In this section we will study the same problem but for a CFT on $R^{1,1}$. Then the relevant problem is to solve  the free field equation for a massive scalar field in Poincar\'{e} AdS. 

\subsection{Lorentzian Solutions}

The metric for the Poincar\'e patch of Lorentzian AdS$_{2+1}$ is given by
\begin{equation}
\mathrm{d}s^2=\frac{1}{z^2}\left(\mathrm{d}t^2+\text{d}z^2+\text{d}x^2\right) \label{Pmetric}
\end{equation}
with the asymptotic boundary at $z = 0$.
In this background the Klein-Gordon equation is given by
\begin{equation}
\left(\partial_z^2 -\frac{1}{z}\partial_z-\partial _t^2 +\partial _x^2-\frac{m^2}{z^2}\right)\Phi(t,z,x) =0. \label{eq:LorentzianKG}
\end{equation}
Substituting the ansatz
\begin{equation}
\Phi \left(t,z,x\right) =\mathrm{e}^{-i\omega t+ikx} f_{\omega  k}(z)
\end{equation}
we get
\begin{equation}
f_{\omega k}''(z) -\frac{1}{z}f_{\omega k}'(z) +\left(\omega^ 2 -k^2-\frac{m^2}{z^2} \right)f_{\omega k}(z)=0. \label{radialODE}
\end{equation}
To solve this ODE we need to consider the cases $-\omega^2 + k^2>0$  (spacelike modes) and $-\omega^2+k^2\leq0$ (timelike modes).

\subsubsection{Timelike Modes}

For timelike modes
\begin{equation}
-\omega ^2+k^2=-q^2\leq 0.
\end{equation}
The two linearly independent solutions to the $z$--ODE are
\begin{subequations}
\begin{align}
f_1(z)=zJ_l(q z)\\
f_2(z)=zY_l(q z)
\end{align}
\end{subequations}
where $J_l$ and $Y_l$ are Bessel functions of the first and second kind respectively, and
$l=\sqrt{1+m^2}\, \in \{0,1,2, \dots \}, \ 
q^2=\omega ^2-k^2$. 
The boundary behaviour of these solutions is
\begin{subequations}
\begin{alignat}{4}
z J_l(q z) &\xrightarrow[z\to0]{}&&  z^{1+l}\left(\frac{q^l}{2^l\Gamma (l)}-\dots \right)\hfill&& \text{\small normalisable\normalsize}\\
z Y_l(q z) &\xrightarrow[z\to0]{}&& z^{1-l}\bigg(\frac{-2^l\Gamma (l)}{q^l\pi }+\text{...} \nonumber \\
&&&\phantom{z^{1-l} (}+z^{2l}\frac{(-1)^lq^l\Gamma (-l)}{2^l\pi }+\text{...}\bigg) \qquad  && \text{\small non--normalisable.\qquad \normalsize}
\end{alignat}
\end{subequations}
As $z \to \infty$,
\begin{subequations}
\begin{alignat}{4}
z J_l(q z)&\xrightarrow[z\to\infty]{} && z^{1/2}\sin \left(\frac{\pi }{4}-\frac{\text{l$\pi $}}{2}+q z\right)\sqrt{\frac{2}{\pi  q}} && \phantom{\text{\small non--normalisable. \normalsize}}\nonumber \\
&&&+z^{-1/2}\sin \left(\frac{\pi}{4}+\frac{\text{l$\pi $}}{2}-q z\right)\frac{\left(4l^2-1\right)}{4\sqrt{2\pi  q^3}}+\text{...}\\
zY_l(q z)&\xrightarrow[z\to\infty]{} && -z^{1/2}\sin \left(\frac{\pi }{4}+\frac{\text{l$\pi $}}{2}-q z\right)\sqrt{\frac{2}{\pi  q}} \nonumber \\
&&&-z^{-1/2}\cos \left(\frac{\pi}{4}+\frac{\text{l$\pi $}}{2}-q z\right)\frac{\left(4l^2-1\right)}{4\sqrt{2\pi  q^3}}+\text{...}
\end{alignat}
\end{subequations}
From these expressions we observe that there are no individual timelike modes that remain finite in the bulk.
Therefore, any solution that is finite must be constructed by integrating over infinitely many such modes.

\subsubsection{Spacelike Modes}

For spacelike modes
\begin{equation}
-\omega^2 + k^2 =q^2 \geq 0.
\end{equation}
The two linearly independent solutions to the $z$--ODE become
\begin{subequations}
\begin{align}
f_1(z)=z I_l\left(q_{\delta }z\right)\\
f_2(z)=zK_l\left(q_{\delta }z\right) \label{Poincare_non-normalisable}
\end{align}
\end{subequations}
where $I_l$ and $K_l$ are modified Bessel functions of the first and second kind respectively, $l$ is as defined above and
$q_{\delta }=\left(-\omega ^2+k^2-\text{i$\delta $}\right)^{1/2}$, with $\delta>0$ an infinitesimal parameter. 
Looking again at the near boundary behaviour of the solutions we find
\begin{subequations}
\begin{align}
z I_l(q z) \xrightarrow[z\to0]{}   z^{1+l}\left(\frac{q^l}{2^l\Gamma (l)}+\frac{q^{2+l}z^2}{2^{2+l}(1+l)\Gamma (1+l)}+O\left(z^3\right)\right) \nonumber \\
\hfill \text{\small normalisable}
\end{align}
\end{subequations}
\begin{subequations}
\begin{alignat}{4}
zK_l(q z)&\xrightarrow[z\to0]{} && z^{1-l}\left(\frac{2^{l-1}\Gamma (l)}{q^l}-\frac{2^{l-3}\Gamma (l)z^2}{q^{l-2}(l-1)}+O\left(z^3\right)\right) \nonumber \\
&&& +z^{1+l}\left(\frac{q^l\Gamma(-l)}{2^{l+1}}+\frac{q^{l+2}z^2\Gamma (-l)}{2^{l+3}(1+l)}+O\left(z^3\right)\right) \nonumber \\
&&&&\text{\small non--normalisable}.\tag{53b} 
\end{alignat}
\end{subequations}
As $z \to \infty$,
{\addtocounter{equation}{-1}
\begin{subequations}
\begin{alignat}{3}
zI_l(q z)&\xrightarrow[z\to\infty]{}&& \frac{z^{1/2}}{\sqrt{2\pi  q}}\left[ \mathrm{e}^{qz}\left(1+\mathrm{O}\left(z^{-1}\right)\right)+\mathrm{e}^{-qz}\left(i(-1)^l+\mathrm{O}\left(z^{-1}\right)\right)\right]\\
z K_l(q z)&\xrightarrow[z\to \infty]{}&& z^{1/2}\mathrm{e}^{-qz}\left[\sqrt{\frac{\pi }{2q }}+\frac{4l^2-1}{8z}\sqrt{\frac{\pi }{2q^3}}+O\left(z^{-2} \right)\right].
\end{alignat}
\end{subequations}}
Here one set of modes, namely the non-normalisable $z K_l(q\,z)$ modes, remain finite at the interior whereas the normalisable ones diverge. 
Consequently, the only physical spacelike modes are the non-normalisable ones.

We are now in position to construct the Lorentzian solutions using the physical modes we have found.
Our choice of boundary conditions for the Lorentzian manifolds dictates that there are no sources present.
Accordingly, we construct Lorentzian solutions using only normalisable modes,
\begin{equation}
\Phi _L\left(t,z,x\right)= \int_{-\infty}^{\infty} \! \frac{\d k}{2\pi} \int_{-\infty}^\infty \! \frac{\d \omega}{2\pi}\Big[
a_{\omega k}\,\mathrm{e}^{-i\omega t+i kx }z\, \theta \left(\omega ^2-k^2\right)
J_l\left(\sqrt{\omega^2-k^2} \, z\right)+\text{c.c.}\Big]. 
\end{equation}

\subsection{Euclidean Solution}

The metric for the Poincar\'e patch of Euclidean AdS$_{2+1}$ can be obtained from the Lorentzi\-an one, (\ref{Pmetric}), by Wick rotating $t= -i\tau$. 
Similarly, the solutions to the Klein--Gordon equation for Euclidean signature can be obtained by analytically continuing the Lorentzian modes and possibly deforming the $\omega$ contour when necessary.
Applying this logic, one finds that, for $\tau\leq 0$, the normalisable Euclidean solution can be cast in the general form
\begin{equation}
\Phi _{E}^- \left(\tau ,z ,x\right)= \int_{-\infty}^{\infty}  \! \frac{\d k}{2\pi}\int _{0}^{\infty}\! \frac{\d\omega}{2\pi i}\, \bigg[ d_{\omega k}\, \mathrm{e}^{\omega \tau +i kx}z\,\theta \left(\omega ^2-k^2\right) 
J_l\left(\sqrt{\omega ^2-k^2}\;z\right) \bigg] \label{n1}\tag{56a}
\end{equation}
and for $\tau \geq 0$ 
\begin{equation}
\Phi _{E}^+ \left(\tau ,z ,x\right)=\int_{-\infty}^{\infty}  \frac{\d k}{2\pi}\int _{0}^{\infty}\! \frac{\d \omega}{2\pi i}\, \bigg[ \tilde{d}_{\omega k}\,\mathrm{e}^{-\omega \tau +i kx}z\,\theta \left(\omega ^2-k^2\right) 
J_l\left(\sqrt{\omega ^2-k^2}\;z\right) \bigg]. \label{n2} \tag{56b}
\end{equation}

The non--normalisable Euclidean solution is constructed using modes proportional to  \(zK_l(p\,z)\), where now $p = (\omega^2+k^2)^{1/2}$.
These are non--normalisable, source modes which we normalise such that, as $z\to 0$, 
\addtocounter{equation}{+1}
\begin{equation}C_{\omega k} zK_l(p\, z) =1 \cdot z^{1-l}+ \ldots\end{equation}
The resulting modes are convoluted with the modes of a source with a $\delta$--function profile, localised in spacetime on the boundary.  We consider a delta function source localised at $\tau = -\epsilon, x=0$, where $\epsilon>0$. Then 
the corresponding bulk solution is given by
\begin{align}
\Phi_E^-\left(\tau,z,x\right)=&\frac{z}{\Gamma (l)2^{l-1}}\int_{-\infty}^{\infty} \frac{\d k}{2\pi}\int_{-\infty}^{\infty} \frac{\d\omega}{2\pi}\, \bigg[ \mathrm{e}^{i \omega \tau +ik x} \phi_{(0)}^-(\omega,k)  
 \left(\omega^2+k^2\right)^{l/2} K_l  \!\left(\!\sqrt{\omega^2+k^2} \,z\right)\!\! \bigg] \nonumber \\[10pt]
\phi_{(0)}^- (\omega, k) =&\, \mathrm{e}^{i\omega \epsilon}. \label{nn1}
\end{align}
Indeed, it is easy to see that in the limit $z \to 0$ this is $\delta$--function source localised at $(\tau, x) = (-\epsilon,0)$.
Similarly, for $\tau \geq 0$ and for a source localised at $(\tau,x) = (\epsilon,0)$, the solution takes the form
\begin{align}
\Phi_E^+\left(\tau,z,x\right)=&\frac{z}{\Gamma (l)2^{l-1}}\int_{-\infty}^{\infty} \frac{\d k}{2\pi}\int_{-\infty}^{\infty}  \frac{\d\omega}{2\pi}\bigg[ \phi_{(0)}^+(\omega,k)\, \mathrm{e}^{i \omega \tau+ik x} 
\left(\omega^2+k^2\right)^{l/2}  K_l  \!\left(\!\sqrt{\omega^2+k^2} \,z\right) \bigg] \nonumber \\[10pt]
\phi_{(0)}^+ (\omega,k) =& \mathrm{e}^{-i\omega \epsilon}. \label{nn2}
\end{align}

\subsection{Matching Conditions} \label{sec:MatchingPoincare}

We will consider the in--in field theory contour and corresponding manifold discussed in section~\ref{contour_discussion} and shown in figure \ref{fig:inincontour}.
Thus, the contour--integrated action and matching conditions are identical to those used for global AdS$_{2+1}$.
The solutions in each manifold, which are constructed by appropriate modifications of the general solutions obtained above, are
\begin{subequations}
\begin{flalign}
0\leq t_1 \leq T: \nonumber \\
 \Phi_L^1(t_1,z,x) =& \int_{-\infty}^{\infty} \! \frac{\d k}{2\pi} \int_{-\infty}^\infty \! \frac{\d \omega}{2\pi} \Big[a_{\omega k}\,\mathrm{e}^{-i\omega t_1+i kx }z\, \theta \left(\omega ^2-k^2\right)J_l\left(\sqrt{\omega^2-k^2} \, z\right)
 \nonumber \\
&\qquad  \qquad \qquad \qquad  +\text{c.c.}\Big], \\[10pt]
T \leq t_2 \leq 2T: \nonumber \\
\Phi^2_L\left(t_2,z,x\right)= &\int_{-\infty}^{\infty} \! \frac{\d k}{2\pi} \int_{-\infty}^\infty \! \frac{\d\omega}{2\pi}\Big[\tilde{a}_{\omega k}\,\mathrm{e}^{-i\omega t_2+i kx }z\, \theta \left(\omega ^2-k^2\right)J_l\left(\sqrt{\omega
^2-k^2} \, z\right) \nonumber \\
&\qquad  \qquad \qquad \qquad +\text{c.c.}\Big],
\end{flalign}
\end{subequations}
for the two Lorentzian segments, and
\begin{subequations}
\begin{flalign}
 -\infty < \tau_0 \leq 0: \nonumber \\
\Phi_E^-\left(\tau_0,z,x\right)&= \quad \nonumber \\
=\frac{z}{\Gamma (l)2^{l-1}}&\int_{-\infty}^{\infty} \frac{\d k}{2\pi}\int_{-\infty}^{\infty} \frac{\d \omega}{2\pi} \bigg[\mathrm{e}^{i \omega (\tau_0+\epsilon) +i
k x} \! \left(\omega^2+k^2\right)^{l/2}\! K_l  \!\left(\!\sqrt{\omega^2+k^2} \,z\right)  \nonumber \\[5pt]
\phantom{=}& \left. + \int_{-\infty}^{\infty}  \frac{\d k}{2\pi}\int _{0}^{\infty}\! \frac{\d \omega}{2\pi i} \, d_{\omega k} \, \mathrm{e}^{\omega \tau_0 +i kx}z\,\theta \left(\omega ^2-k^2\right)J_l\left(\sqrt{\omega ^2-k^2}\;z\right)\right],  &\label{PEu1} \\[10pt]
 0\leq \tau_3 <\infty: \nonumber\\
\Phi_E^+\left(\tau_3,z,x\right) &=  \nonumber \\
= \frac{z}{\Gamma (l)2^{l-1}}&\int_{-\infty}^{\infty} \frac{\d k}{2\pi}\int_{-\infty}^{\infty} \frac{\d \omega}{2\pi} \bigg[  \mathrm{e}^{i \omega (\tau_3-\epsilon)+ik x}\, \left(\omega^2+k^2\right)^{l/2} K_l  \!\left(\!\sqrt{\omega^2+k^2} \,z\right)  \nonumber \\[5pt]
\phantom{=}&\left.  +\int_{-\infty}^{\infty}  \frac{dk}{2\pi}\int _{0}^{\infty}\! \frac{d\omega}{2\pi i}\tilde{d}_{\omega k}\, \mathrm{e}^{-\omega \tau_3 +i kx}z\,\theta \left(\omega ^2-k^2\right)J_l\left(\sqrt{\omega ^2-k^2}\;z\right) \right],& \label{PEu2} 
\end{flalign}
\end{subequations}
for the two Euclidean segments.
The Lorentzian solutions are purely normalisable whereas the Euclidean solutions are linear combinations of a non--normalisable piece and a normalisable piece.
 In momentum space we saw that the individual modes are either Bessel functions of the first kind, $J_l$, or modified Bessel functions of the second kind, $K_l$.
These functions are not orthogonal to each other.
We circumvent this complication by making use of the following two integrals of Bessel functions~\cite{jeffrey2007table}
\begin{subequations}
\begin{align}
\int_0^\infty dz \; z J_n(za) J_n(zb) = \frac{1}{a} \delta(b-a), \qquad &a,b \in \mathbb{R} \label{Bessel 1}\\
\int_0^\infty dz \; zK_\nu (za) J_\nu(zb) = \frac{b^\nu}{a^\nu (a^2+b^2)}, \qquad &\text{Re}(a)>0, \; b>0 \label{Bessel 2}.
\end{align}
\end{subequations}

To extract individual modes from our solutions we perform the following steps.
Given a field $\Phi(t,z,x)$ or its time derivative $\partial_t \Phi(t,z,x)$, where $t$ here can be either real or imaginary time, we multiply by 
$\theta \lc \omega^2 - k^2\right)$ $J_l\left(\sqrt{\omega^2 -k^2}\, z\right) \mathrm{e}^{-i kx}$ and integrate first over $x$ from $-\infty$ to $+\infty$ and then over $z$ from zero to $+\infty$,
\begin{equation}
\int_0^\infty \mathrm{d}z \, \theta(\omega^2 - k^2) J_l\left( \sqrt{\omega^2-k^2} \, z \right) \int_{-\infty}^\infty d x \, \mathrm{e}^{-ikx} \Phi(t,z,x )\big|_{\text{on matching surface}}. \label{extraction sequence}
\end{equation}
To perform the $z$ integral one needs to use either equation (\ref{Bessel 1}) or (\ref{Bessel 2}).
The Heaviside step function is to ensure that the conditions associated with these two equations are satisfied.
Some of  the details of this calculation are given in appendix~\ref{sec:appendix}.
 
 Applying the matching conditions to these solutions and using the above prescription to extract individual modes we finally obtain the following relations which hold for $\omega^2>k^2$. 
 Note that normalisable modes exist only for $\omega^2>k^2$ so the above matching conditions are sufficient for our purposes.

From the matching conditions at $\tau_0=0,\, t_1 = 0$, between the past Euclidean cap and the first Lorentzian manifold, we obtain
\begin{subequations}
\begin{align}
 a_{|\omega| k}+a_{-|\omega| \,-k}^{\dagger} =& \frac{\left(\omega^2 -k^2\right)^{l/2}\pi}{\Gamma(l) 2^{l-1}} \mathrm{e}^{-|\omega| \epsilon} \nonumber \\
 =&\frac{\left( \omega^2 - k^2 \right)^{l/2} \pi}{\Gamma(l) 2^{l-1} } \phi_{(0)}^-(i |\omega| , k) \\
a_{-|\omega| k}+ a_{|\omega| \, -k}^\dagger=& -i d_{|\omega| k}.
\end{align}
\end{subequations}
 From the matching conditions at $t_1 = T, \, t_2 = T$, between the two Lorentzian manifolds,
\begin{subequations}
\begin{align}
 a_{|\omega| k}+a_{-|\omega| \,-k}^{\dagger} =&\left(\tilde{a}_{-|\omega|k} + \tilde{a}_{|\omega| -k}^\dagger\right) \mathrm{e}^{2i|\omega|T} \\
a_{-|\omega|k} + a_{|\omega| -k}^\dagger =& \left( \tilde{a}_{|\omega|k} + \tilde{a}_{-|\omega| -k}^\dagger \right) \mathrm{e}^{-2i|\omega| T}
\end{align}
\end{subequations}
Finally, the matching conditions at $t_2 = 2T, \, \tau_3 = 0$, between the second Lorentzian manifold and the future Euclidean cap give
\begin{subequations}
\begin{align}
 \tilde{a}_{|\omega| k}+\tilde{a}_{-|\omega| \, -k}^{\dagger} =&  \frac{\left(\omega^2 -k^2\right)^{l/2}\pi}{\Gamma(l) 2^{l-1}} \mathrm{e}^{-|\omega| (\epsilon-2iT)} \nonumber \\
  =&\frac{\left(\omega^2 -k^2\right)^{l/2}\pi}{\Gamma(l) 2^{l-1}} \mathrm{e}^{2i |\omega|T} \phi_{(0)}^+(-i|\omega|,k) \\
 \tilde{a}_{-|\omega| k}+ \tilde{a}_{|\omega| \, -k}^{\dagger} =&-i \tilde{d}_{|\omega|k} \mathrm{e}^{-2i |\omega| T}. 
 \end{align}
 \end{subequations}
Given the matching relations it is easier to redefine the Lorentzian coefficients by introducing $b_{\omega k} = a_{|\omega| k} + a_{-|\omega| -k}^\dagger$ and $b_{\omega -k}^\dagger = a_{-|\omega| k} + a_{|\omega| -k}^\dagger$ for the first Lorentzian manifold and $\tilde{b}_{\omega k} =\tilde{a}_{|\omega| k} + \tilde{a}_{-|\omega| -k}^\dagger$ and $\tilde{b}_{\omega -k}^\dagger = \tilde{a}_{-|\omega| k} + \tilde{a}_{|\omega| -k}^\dagger$ for the second Lorentzian manifold.
In terms of these new coefficients the solutions become
\begin{align}
\Phi_L^1(t_1,z,x) = \int_{-\infty}^{\infty} \! \frac{\d k}{2\pi} \int_0^\infty \! \frac{\d\omega}{2\pi} &\bigg[ \left(b_{\omega k}\,\mathrm{e}^{-i\omega t_1+i kx} + b_{\omega -k}^\dagger \mathrm{e}^{i\omega t_1 +ikx} \right) \nonumber \\
& \; z\, \theta \left(\omega ^2-k^2\right)J_l\left(\sqrt{\omega
^2-k^2} \, z\right) \bigg],
\end{align}
with an analogous expression for $\Phi_L^2(t_2,z,x)$.

Re--expressing the matching conditions in terms of $b$'s and $\tilde{b}$'s,
\begin{subequations}
\begin{align}
b_{\omega k} =& \frac{\left( \omega^2 - k^2 \right)^{l/2} \pi}{\Gamma(l) 2^{l-1} } \phi_{(0)}^-(i \omega , k)=\tilde{b}_{\omega -k}^\dagger \mathrm{e}^{2i \omega T}=-i \tilde{d}_{\omega -k}^\dagger  \\
b_{\omega -k}^\dagger=& -i d_{\omega k}=\tilde{b}_{\omega k} \mathrm{e}^{-2i\omega T} =\frac{\left(\omega^2 -k^2\right)^{l/2}\pi}{\Gamma(l) 2^{l-1}} \phi_{(0)}^+(-i\omega,k)
 \end{align}
 \end{subequations}
where the frequency $\omega$ is greater or equal to zero.
Note that had we not chosen the source insertion points in the past and future Euclidean caps to be the same, reality conditions for the Lorentzian solutions would dictate that they have to be the same.

Identifying the coefficients of $\mathrm{e}^{-i\omega t}$ ($\mathrm{e}^{-\omega \tau}$) as the positive frequency oscillatory (exponential) modes and the coefficients of $\mathrm{e}^{+i\omega t}$ as the negative ones, we see that our modes evolve in an analogous way as we saw in the global case. 
In particular, the positive frequency normalisable modes in the first Lorentzian manifold are sourced by exponentially decaying positive frequency source modes in the past Euclidean manifold whereas the positive frequency source modes decay.
The positive frequency Lorentzian modes from the first manifold then evolve across the matching surface at $t_1=T=t_2$ to become negative frequency modes in the second Lorentzian manifold and finally they become negative frequency normalisable modes in the future Euclidean manifold. 
There are no positive frequency normalisable modes in the future manifold as these grow exponentially as $\tau_3\to \infty$.

The negative frequency normalisable modes in the first Lorentzian manifold are the evolution of positive frequency normalisable modes which we have included in the past Euclidean manifold.
As they evolve across the matching surface into the second Lorentzian manifold they become the positive frequency normalisable modes which are associated to negative frequency source modes turned on in the future Euclidean manifold. 

Returning to the Lorentzian fields, we can now replace the arbitrary coefficients $b_{\omega k}$ and $\tilde{b}_{\omega k}$ with the above results to obtain
\begin{subequations}
\begin{alignat}{3}
\Phi_L^1(t_1,z,x) =& \frac{z}{\Gamma(l) 2^l}\! &\int_{-\infty}^{\infty} \frac{\d k}{2\pi} \int_0^\infty \! \d\omega \bigg[ \!\! \left( \phi_{(0)}^-(i\omega,k)\mathrm{e}^{-i\omega t_1} + \phi_{(0)}^+(-i\omega,k) \mathrm{e}^{i\omega t_1}\right)  \nonumber \\
&&\mathrm{e}^{ikx}\left(\omega^2 - k^2\right)^{l/2}  \theta \left(\omega ^2-k^2\right)J_l\left(\sqrt{\omega
^2-k^2} \, z\right)\!\! \bigg], \label{Psol1} \\[10pt]
\Phi_L^2(t_2,z,x) =& \frac{z}{\Gamma(l) 2^l} & \int_{-\infty}^{\infty}   \frac{\d k}{2\pi} \int_0^\infty \! \d \omega\bigg[ \!\! \left( \phi_{(0)}^+(-i\omega,k)\mathrm{e}^{-i\omega t_2} + \phi_{(0)}^-(i\omega,k) \mathrm{e}^{i\omega t_2}\right) \nonumber \\
&& \mathrm{e}^{ikx}\left(\omega^2 - k^2\right)^{l/2}  \theta \left(\omega ^2-k^2\right)J_l\left(\sqrt{\omega
^2-k^2} \, z\right)\!\! \bigg].
\label{Psol2}
\end{alignat}
\end{subequations}
Equations (\ref{Psol1}) and (\ref{Psol2}) demonstrate explicitly how the Euclidean source modes generate the purely normalisable solutions in the Lorentzian bulk.
\subsection{1-point  function}

We will now extract the 1-point function to verify that the solution indeed describes an excited state. For this we need to extract 
the coefficient $\phi_{(2 \Delta -2)}$, which in our case is the leading order coefficient of the bulk solution. As in the case of global AdS, we consider the case where the operator is in the upper part of the contour so the relevant field is $\Phi_L^1$. Then 
\begin{align}
\phi_{(2 \Delta-2)}(t,x) =&  \lim_{z\to0}z^{\Delta} \Phi_L^1(z,t,x) 
 =\frac{1}{ 2^{2l-1} \Gamma(l) \Gamma(l+1)} \\
& \int_{-\infty}^{\infty} \! \frac{\d k}{2\pi} \int_{0}^\infty \! \d\omega\, 
\bigg[ \theta\! \left( \omega^2 - k^2 \right) \left(\omega^2 - k^2 \right)^{l}  \mathrm{e}^{-\omega \epsilon+ikx} \cos(\omega t ) \bigg] \nonumber
\end{align}
 Eliminating first the Heaviside step function and setting $\omega =rk$, we obtain
\begin{align}
\phi_{(2 \Delta-2)}(t,x)  =\frac{1}{ 2^{2l-1} \Gamma(l) \Gamma(l+1)} &\int_{0}^{\infty}  \! \frac{\d k}{2\pi} \int_{1}^\infty \! \d r\Big[ k^{2l+1} \left(r^2 - 1 \right)^{l} \mathrm{e}^{-kr \epsilon } \\
&\quad \lc \cos\big(k(rt+x)\big)+\cos\big(k(rt - x)\big)\rc\Big]. \nonumber 
\end{align}
Then we perform the $k$ integral,
\begin{align}
 \phi_{(2 \Delta-2)}&(t,x)= \frac{(-1)^{\Delta} \Gamma(2\Delta)}{ 2^{2\Delta-1}\pi \Gamma(\Delta-1) \Gamma(\Delta)} \int_{1}^\infty \! \d r  \,  \bigg[ \left(r(t+i\epsilon)-x\right)^{-2\Delta} +  \\
 &\left(r(t+i\epsilon)+x\right)^{-2\Delta}+\left(r(t_1-i\epsilon)+x\right)^{-2\Delta} +\left(r(t-i\epsilon)-x\right)^{-2\Delta}\bigg] \left(r^2 - 1 \right)^{\Delta-1}, \nonumber
\end{align}
and finally, we compute the $r$ integral,
\begin{equation}
 \phi_{(2 \Delta-2)}(t,x) =  -\frac{l}{\pi}\left(\frac{1}{ \left(-(t-i\epsilon)^2 +x^2\right)^{\Delta}}+\frac{1}{ \left(-(t+i\epsilon)^2 + x^2\right)^{\Delta}}\right)
 \end{equation}
 and thus,
 \begin{equation}
\langle \mathcal{O}_\Delta(t,x)\rangle = \frac{2 l^2}{\pi}\left(\frac{1}{ \left(-(t-i\epsilon)^2  +x^2\right)^{\Delta}}+\frac{1}{ \left(-(t+i\epsilon)^2 + x^2\right)^{\Delta}}\right)  \label{1ptfunctionPoincare}
\end{equation}
This is indeed equal to value we got via a QFT computation in (\ref{R11}). In our case, $\tilde{C}=2 l^2/\pi$, which  is the standard supergravity normalisation of the 2-point function. Note also that the normalisations in (\ref{1ptfunctionGlobal}) and (\ref{1ptfunctionPoincare}) are related as in the footnote \ref{normalization}, as they should.

\section{Discussion} \label{sec:discussion}

We presented in this paper a construction of a bulk solution dual to a general excited CFT state, $| \Delta \rangle$, where $\Delta$ is the scaling dimension. By the operator-state correspondence, the state is generated by an operator ${\cal O}_\Delta$ acting on the vacuum. The corresponding bulk solution at linearised level involves only the bulk scalar $\Phi$ which is dual to the operator ${\cal O}_\Delta$. This part is universal: it is the same for all CFTs whose spectrum contains an operator with such dimension. 
To construct the full bulk solution we need more information about the CFT. In particular, we need to know the OPE of ${\cal O}_\Delta$  with itself. All bulk fields that are dual to operators that appear in this OPE are necessarily turned on in the bulk. 

In this paper we discussed in detail the construction of the universal part, for states of two dimensional CFTs either on $R \times S^1$ or $R^{1,1}$.
From the bulk perspective this leads to the construction of solutions of free scalar field equations  either in global $AdS_3$ or Poincar\'{e} $AdS_3$. 
The solutions describe normalisable modes and their coefficients are directly related to the dual state. In more detail, the CFT state is generated by a Euclidean path integral which contains a source for ${\cal O}_\Delta$ and the coefficients of the bulk normalisable modes are given in terms of the source. 
Normalisable modes
describe bulk local excitations and thus our results give a direct relation between CFT states and bulk excitations. To substantiate the claim that these solutions 
are dual to the state $| \Delta \rangle$, we computed the 1-point function of local operators 
both in the CFT and in the bulk and found perfect agreement\footnote{As emphasised in section \ref{QFT}, this agreement is a non-trivial check that we are constructing the correct path integral. To holographically compute expectation values in the state $| \Delta \rangle$ we would need the solution to quadratic order in the bulk fields.}.  Our discussion generalizes straightforwardly to higher dimensions.

To go beyond this leading order computation, one needs to be more specific about the CFT (as mentioned above). In particular, one would need to take into account the backreaction to the metric. Given appropriate CFT data (for a CFT with a known bulk dual),
the construction of the bulk solution  dual to any given state can proceed along the same lines. It would be interesting to explicitly carry this out in detail in concrete examples.

In our discussion we explicitly demonstrated how a solution of the bulk field equations is reconstructed from QFT data: 
given a Schwinger-Keldysh contour and insertions we constructed a unique bulk solution. To make this more explicit 
one may rewrite the bulk solution in the Lorentzian part in the following form,
\begin{equation}
\Phi(t,r, \phi) = \int_{\partial \text{AdS}}\!\!\!\!\!{\d}t' \,{\d \phi'} \; K(t,r,\phi|t',\phi') \langle O(t', \phi') \rangle
\label{eq:smearing}
\end{equation}
where  $K(t,r, \phi| \hat{t}, \hat{\phi})$, is the so-called smearing function, whose detailed form will not be needed here.
%
The derivation of this relation follows closely the discussion in \cite{Hamilton:2006az} and it will not be repeated here.

For us (\ref{eq:smearing}) is a map between expectation values of the boundary theory and classical fields in the bulk.  In \cite{Hamilton:2006az} the idea was different. The main point was to look for CFT operators that behave like bulk local operators.
The initial ansatz in \cite{Hamilton:2006az} was
\begin{equation}
\hat{\Phi}(t,r, \phi) = \int_{\partial \text{AdS}}\!\!\!\!\!{\d}t' \,{\d \phi'} \; K(t,r,\phi|t',\phi') O(t', \phi'),
\label{eq:smearing2}
\end{equation}
and the smearing function $K(t,r,\phi|t',\phi')$ was fixed by rewriting the bulk normalisable modes in this form. The hat on the left hand side indicates that this is a quantum operator. If we quantize canonically  the bulk scalar field then the coefficients $b_{nk}$ and 
$b_{nk}^\dagger$ of the normalisable modes (see (\ref{eq:normalizable})) are promoted to creation and annihilation operators. 
However, the matching condition relates these coefficients to a CFT source and the latter is not a quantum operator. One may still reconcile the two pictures if one considers the bulk solutions as being associated with a coherent state, as was recently argued in
\cite{Botta-Cantcheff:2015sav}.  Then  the eigenvalue of the annihilation operator acting on the coherent state would be equal to the value of the source. This would give a map from states $|\Delta \rangle$ of the CFT to coherent states in the bulk and
it would be interesting to understand this map in more detail. 

As emphasised, (\ref{eq:smearing}) and (\ref{eq:smearing2}) hold at the linearised level in the bulk (free fields)\footnote{This is also the leading term  in the 't Hooft large $N$ limit, if we normalise the CFT operators such that their 2-point function has coefficient 1 in the  large $N$ limit. One should keep in mind however that with this normalisation the subleading terms in $N$ do not necessarily correspond to quantum loops, see the discussion in section \ref{QFT}.}.  While (\ref{eq:smearing}) and (\ref{eq:smearing2}) may be related at this order, it is not clear this will continue to be the case at non-linear level. 
There has been work in extending (\ref{eq:smearing2}) to higher orders, see for example \cite{Kabat:2011rz, Kabat:2012av, Kabat:2012hp, Kabat:2013wga, Kabat:2015swa}. In these papers, the map 
is modified by including additional terms on the RHS  of (\ref{eq:smearing2}),  which are double-trace operators. 
The coefficients are then fixed by requiring bulk locality. In our case, the full bulk solution will instead involve many additional 
bulk fields, which are dual to single-trace operators.  It would be interesting to clarify the relation between the two reconstruction formulae at non-linear order.

Another application of our construction is in the context of the fuzzball program \cite{Mathur:2005zp, Bena:2007kg, Skenderis:2008qn, Balasubramanian:2008da}.  As was argued in \cite{Skenderis:2006ah, Kanitscheider:2006zf, Kanitscheider:2007wq,Skenderis:2008qn}, the fuzzball solutions for black holes with AdS throats are the bulk solutions dual to the states that account for black hole entropy. 
In all previous works, fuzzball solutions were constructed  by solving  supergravity equations and the relation to CFT states was only studied afterwards (for a class of fuzzballs). The construction here allows one to pursue a direct (iterative) construction of bulk solutions dual to individual states. It would be interesting to carry out such computations. One may also use the results here to 
sharpen an old argument \cite{unpublished} that the number of {\it supergravity} solutions dual to the 3-charge BPS black holes cannot exceed that of the 2-charge ones. This will be discussed elsewhere.

\section*{Acknowledgements}
This work was supported by the Science and Technology Facilities Council (Consolidated Grant ``Exploring the Limits of the Standard Model and Beyond'') and by the Engineering and Physical Sciences Research Council. KS thanks the 2015 Simons Center Summer Workshop and the Galileo Galilei Institute for Theoretical Physics for hospitality and the INFN for partial support during the completion of this work. 

\appendix

\section{Matching conditions for the Poincar\'e AdS} \label{sec:appendix}

Here we demostrate how individual modes can be extracted from the solutions obtained for the Poincar\'e patch of AdS.
We only present the calculations for the matching surface at $\tau_0 = 0, t_1 = 0$ but the same method can be applied straightforwardly to the other matching surfaces.

Our analysis makes use of the following two identities of the Bessel functions
\begin{equation}
\int_0^\infty dz \; z J_n(za) J_n(zb) = \frac{1}{a} \delta(b-a) \label{appBessel 1}
\end{equation}
\begin{equation}
\int_0^\infty dz \; zK_\nu (za) J_\nu(zb) = \frac{b^\nu}{a^\nu (a^2+b^2)} \label{appBessel 2}.
\end{equation}
Focusing first on the Lorentzian solution, on the hypersurface located at $t_1=0$ the field and its derivative are given by
\begin{subequations}
\begin{alignat}{3}
\Phi _L^1\left(t_1,z,x\right)\big|_{t_1=0}=&\int_{-\infty}^\infty\! \frac{\d k}{2\pi}&&\int_{-\infty}^{\infty} \frac{\d \omega}{2\pi} \bigg[ \left(a_{\omega k} \mathrm{e}^{ ik x }+a_{\omega k}^*\mathrm{e}^{-ikx} \right)z \nonumber \\
&&& \theta \left(\omega ^2-k^2\right) J_l\left(\sqrt{\omega^2-k^2}\; z\right)\bigg] \\
-i \partial_{t_1} \Phi _L^1\left(t_1,z,x\right)\big|_{t_1=0}=&\int_{-\infty}^\infty\! \frac{\d k}{2\pi} &&\int_{-\infty}^{\infty} \frac{\d \omega}{2\pi} \bigg[ \left(-  a_{\omega k} \mathrm{e}^{ikx} + a_{\omega k}^* \mathrm{e}^{-ikx} \right)\! \omega  z \nonumber \\
&&& \theta \!\left(\omega ^2-k^2\right) J_l\left(\sqrt{\omega^2-k^2} \,z\right)\bigg]. 
\end{alignat}
\end{subequations}
Multiplying the above expressions by $\theta \lc \omega^2 - k^2\right)J_l\left(\sqrt{\omega^2 -k^2}\, z\right) \mathrm{e}^{-i kx}$ and integrating first over $x$ from $-\infty$ to $+\infty$ and then over $z$ from zero to $+\infty$, we find
\begin{subequations}
\begin{align}
\int_0^\infty \!\!\! \d z \,  &\theta \! \left(\omega^2 -k^2\right)J_l(\sqrt{\omega^2-k^2} \, z) \int_{-\infty}^{\infty}\!\!\d x \, \mathrm{e}^{-ikx} \Phi_L^1\left(t_1,z,x\right)\big|_{t_1=0}= \nonumber \\ 
& =\frac{\theta(\omega^2-k^2)}{2\pi |\omega|}\left(a_{|\omega| k}+a_{-|\omega|,k}+a_{|\omega|\, -k}^* + a_{-|\omega| \,-k}^*\right) \\[5pt]
\int_0^\infty \!\!\! \d z \,  &\theta \! \left(\omega^2 -k^2\right)J_l(\sqrt{\omega^2-k^2} \, z) \int_{-\infty}^{\infty}\!\!\d x \,\mathrm{e}^{-i kx} \left(-i \partial_{t_1} \Phi _L^1 \left(t_1,z,x \right)\big|_{t_1=0}\right)= \nonumber \\ 
&=\frac{\theta(\omega^2-k^2)}{2\pi}\left(-a_{|\omega| k}+a_{-|\omega| k} - a_{-|\omega| \,-k}^*+a_{|\omega|\,-k}^*\right)
\end{align}
\end{subequations}
In more details:
\begin{align}
&\int_0^\infty \!\! \d z \,  \theta\left(\omega^2 - k^2\right) J_l(\sqrt{\omega^2-k^2} \, z) \int_{-\infty}^{\infty} \!\! \d x \, \mathrm{e}^{-i kx} \Phi _L^1 \left(t_1,z,x\right)\big|_{t_1=0}= &\nonumber \\ 
&=\int_0^\infty \!\! \d z\int_{-\infty}^\infty\!\! \d x \int_{-\infty}^{\infty} \! \frac{\d k^\prime}{2\pi} \int_{-\infty}^{\infty} \! \frac{\d \omega^\prime}{2\pi} \theta\left(\omega^2 - k^2\right) \theta\!\left(\omega^{\prime 2} - k^{\prime 2} \right) \left[ a_{\omega^\prime k'} \mathrm{e}^{i(k' - k)x} \right. & \nn \\
&\phantom{=} \left. + a_{\omega' k'}^* \mathrm{e}^{-i(k'+k)x} \right] z\,  J_l\left(\sqrt{\omega^2 - k^2} \, z\right) J_l\left( \sqrt{\omega'^2 - k'^2} \, z \right) & \nn \\
&= \int_{-\infty}^{\infty} \!\! \frac{\d \omega'}{2\pi} \int_0^\infty\!\! \d z \, \theta\! \left(\omega
^{\prime2}-k^2\right) \theta \! \left(\omega^2 - k^2\right) \left(a_{\omega' k}+a_{\omega' \, -k}^* \right) z & \nn \\
&\phantom{=} J_l\left(\sqrt{\omega^2 - k^2} \, z\right) J_l\left( \sqrt{\omega'^2 - k^2} \, z \right) & \nn \\
&= \int_{-\infty}^\infty\!\!  \frac{\mathrm{d}\omega'}{2\pi}    \theta\! \left(\omega^{\prime2}-k^2\right)\! \theta \! \left(\omega^2 - k^2\right)\!\! \left(a_{\omega' k}+a_{\omega' \, -k}^* \right) \frac{\delta \left( \sqrt{\omega^{\prime2}-k^2}-\sqrt{ \omega^2 - k^2} \right)}{\sqrt{\omega^2-k^2}}&
\end{align}
where in the last line we used (\ref{appBessel 1}) to perform the $z$ integral.

To proceed we make use of the relation
\begin{equation}
\delta \left( \sqrt{\omega^{\prime2}-k^2}-\sqrt{ \omega^2 - k^2} \right) = \frac{\sqrt{\omega^2-k^2}}{|\omega|} \big[\delta\left(\omega' +|\omega|\right) 
+ \delta \left(\omega' -|\omega|\right) \big] \label{deltafunction} 
\end{equation}
to obtain
\begin{align}
&\int_0^\infty \mathrm{d}z \,\theta\left(\omega^2-k^2\right) J_l(\sqrt{\omega^2-k^2} \, z) \int_{-\infty}^{\infty}\!\! \d x \, \mathrm{e}^{-i kx} \Phi _L^1\left(t_1,z,x\right)\big|_{t_1=0}= &\nonumber \\ 
&= \int_{-\infty}^\infty\!  \frac{\mathrm{d}\omega'}{2\pi |\omega|} \left(a_{\omega' k}+a_{\omega'\, -k}^* \right) \theta\! \left(\omega^2-k^2\right) \theta \! \left(\omega
^{\prime2}-k^2\right)  \Big[ \delta\left(\omega' +|\omega| \right) &\nn \\
&\phantom{=} +\delta \left(\omega' -|\omega|\right) \Big]& \nn \\
&=\frac{\theta(\omega^2-k^2)}{2\pi |\omega|}\left(a_{|\omega| k}+a_{-|\omega| k}+a_{|\omega|\, -k}^* + a_{-|\omega|\,-k}^*\right). &\hfill \Box 
\end{align}
The computation for the derivative is very similar.

Focusing now on the Euclidean solution, on the hypersurface located at $\tau_0 = 0$, the field and its derivative are given by
\begin{subequations}
\begin{align}
\Phi_E^-\left(\tau_0,z,x\right)\big|_{\tau_0=0}=&\frac{z}{\Gamma (l)2^{l-1}}\int_{-\infty}^{\infty} \frac{\d k}{2\pi}\int_{-\infty}^{\infty} \frac{\d \omega}{2\pi}  \Big[ \mathrm{e}^{i \omega \epsilon +i
k x} \, \left(\omega^2+k^2\right)^{l/2}\nn  \\
& K_l  \!\left(\!\sqrt{\omega^2+k^2} \,z\right) \Big] +  \int_{-\infty}^{\infty}  \frac{\d k}{2\pi}\int _{0}^{\infty}\! \frac{\d \omega}{2\pi i} \, \Big[ b_{\omega k} \, \mathrm{e}^{i kx}z \nn \\
&\theta \left(\omega ^2-k^2\right) J_l\left(\sqrt{\omega ^2-k^2}\;z\right) \Big]  \\[10pt]
\partial_{\tau_0} \Phi_E^- \left(t_0,z,x\right)\big|_{\tau_0=0}=&\frac{z}{\Gamma (l)2^{l-1}}\int_{-\infty}^{\infty} \frac{\d k}{2\pi}\int_{-\infty}^{\infty} \frac{\d \omega}{2\pi}  \, \Big[ i\,   \omega \mathrm{e}^{i \omega \epsilon +i
k x} \, \left(\omega^2+k^2\right)^{l/2}\nn \\
& K_l  \!\left(\!\sqrt{\omega^2+k^2} \,z\right)\Big]+\int_{-\infty}^{\infty}  \frac{\d k}{2\pi}\int _{0}^{\infty}\! \frac{\d \omega}{2\pi i} \, \Big[ \omega \, b_{\omega k} \, \mathrm{e}^{i kx}z\nn \\
&\theta \left(\omega ^2-k^2\right) J_l\left(\sqrt{\omega ^2-k^2}\;z\right) \Big] 
\end{align}
\end{subequations}
By using the same method we find
\begin{subequations}
\begin{align}
&\int_0^\infty \!\!\! \d z \,\theta\left(\omega^2 -k^2\right)J_l(\sqrt{\omega^2-k^2} \, z) \int_{-\infty}^{\infty}\!\! \d x \, \mathrm{e}^{-ikx} \Phi_E^-\left(\tau_0,z,x\right)\big|_{\tau_0=0}= &\nonumber \\ 
&=\theta \!\left( \omega^2 -k^2 \right) \left( \frac{\left(\omega^2 - k^2 \right)^{l/2}}{2^l \Gamma(l) |\omega|} \mathrm{e}^{-|\omega| \epsilon} + \frac{d_{|\omega| k}}{2\pi i |\omega|} \right)& \\[10pt]
&\int_0^\infty \!\!\! \d z \, \theta\! \left(\omega^2 - k^2\right) J_l\lc \sqrt{\omega^2-k^2} \, z \rc \int_{-\infty}^{\infty}\!\! \d x \, \mathrm{e}^{-i kx} \left( \partial_{\tau_0} \Phi _E^- \left(\tau_0,z,x \right)\big|_{\tau_0=0}\right)= &\nonumber \\ 
&=\theta\!\left(\omega^2-k^2\right)\left(-\frac{\left(\omega^2 - k^2\right)^{l/2}}{2^l \Gamma(l)} \mathrm{e}^{-|\omega| \epsilon} + \frac{d_{|\omega| k} }{2\pi i} \right).&
\end{align}
\end{subequations}
Obtaining these results requires a bit of extra work because our Euclidean solutions consists of two terms, one of which is in terms of the modified Bessel function of the second kind and therefore we need to use (\ref{appBessel 2}) and perform a contour integration in the $\omega$ plane.

 In more detail, this is done as follows,
\begin{align}
&\int_0^\infty \!\! \d z \,  \theta\left(\omega^2 - k^2\right) J_l(\sqrt{\omega^2-k^2} \, z) \int_{-\infty}^{\infty} \!\! \d x \, \mathrm{e}^{-i kx} \Phi _E^- \left(\tau_0,z,x\right)\big|_{\tau_0=0}=& \nonumber \\
&=\int_0^\infty \!\!\! \d z \! \int_{-\infty}^\infty\!\! \! \d x \! \int_{-\infty}^{\infty} \! \! \frac{\d k'}{2\pi} \int_{-\infty}^{\infty} \! \! \frac{\d \omega'}{2\pi} \Bigg[ \frac{z \, \theta \! \left(\omega^2 - k^2\right)\left(\omega'^2+k'^2\right)^{l/2}\mathrm{e}^{i \omega' \epsilon - i (k-k') x}}{2^{l-1} \Gamma(l)}  & \nn \\
&\phantom{=} J_l\left( \sqrt{ \omega^2 -k^2} \, z\right) K_l\left(\sqrt{\omega'^2+k'^2} \, z\right)\Bigg] +\int_0^\infty \!\!\! \d z \! \int_{-\infty}^\infty\!\! \! \d x \! \int_{-\infty}^{\infty} \! \! \frac{\d k'}{2\pi} \int_{0}^{\infty} \! \! \frac{\d \omega'}{2\pi i}\Bigg[ \nn & \\
&\phantom{=}d_{\omega' k'} \mathrm{e}^{i(k-k')x}  \theta \!\left( \omega^2-k^2\right) \theta\!\left( \omega'^2 - k'^2\right)
zJ_l\left(\sqrt{\omega^2-k^2}\,z\right)J_l\left(\sqrt{\omega'^2-k'^2}\,z\right) \Bigg] &\nn \\
&=I_1 +I_2& 
\end{align}
where
\begin{subequations}
\begin{align}
I_1=&\int_0^\infty \!\!\! \d z \! \int_{-\infty}^\infty\!\! \! \d x \! \int_{-\infty}^{\infty} \! \! \frac{\d k'}{2\pi} \int_{-\infty}^{\infty} \! \! \frac{\d \omega'}{2\pi} \Bigg[ \frac{z \, \theta \! \left(\omega^2 - k^2\right)\left(\omega'^2+k'^2\right)^{l/2}\mathrm{e}^{i \omega' \epsilon - i (k-k') x}}{2^{l-1} \Gamma(l)}  \nn \\
& J_l\left( \sqrt{ \omega^2 -k^2} \, z\right) K_l\left(\sqrt{\omega'^2+k'^2} \, z\right)\Bigg], \\[10pt]
I_2 =& \int_0^\infty \!\!\! \d z \! \int_{-\infty}^\infty\!\! \! \d x \! \int_{-\infty}^{\infty} \! \! \frac{\d k'}{2\pi} \int_{0}^{\infty} \! \! \frac{\d \omega'}{2\pi i}\Bigg[ d_{\omega' k'} \mathrm{e}^{i(k-k')x}  \theta \!\left( \omega^2-k^2\right) \theta\!\left( \omega'^2 - k'^2\right) \nn \\
& zJ_l\left(\sqrt{\omega^2-k^2}\,z\right)J_l\left(\sqrt{\omega'^2-k'^2}\,z\right) \Bigg]. 
\end{align}
\end{subequations}
The computation of $I_2$ is identical to what we did for the Lorentzian field above,
\begin{align}
I_2 =& \int_0^\infty \!\!\! \d z \! \int_{-\infty}^\infty\!\! \! \d x \! \int_{-\infty}^{\infty} \! \! \frac{\d k'}{2\pi} \int_{0}^{\infty} \! \! \frac{\d \omega'}{2\pi i}\, \Bigg[ z \, d_{\omega' k'} \mathrm{e}^{i(k-k')x}  \theta \!\left( \omega^2-k^2\right) \theta\!\left( \omega'^2 - k'^2\right)\nn \\
\phantom{I_2 =} &J_l\left(\sqrt{\omega^2-k^2}\,z\right)J_l\left(\sqrt{\omega'^2-k'^2}\,z\right) \Bigg] \nn \\
\phantom{I_2}=&\int_0^\infty \!\!\! \d z \int_{0}^{\infty} \! \! \frac{\d \omega'}{2\pi i}\, \Bigg[ z d_{\omega' k}  \theta \!\left( \omega^2-k^2\right) \theta\!\left( \omega'^2 - k^2\right)J_l\left(\sqrt{\omega^2-k^2}\,z\right)\nn \\
 \phantom{I_2 =}& J_l\left(\sqrt{\omega'^2-k^2}\,z\right) \Bigg] \nn \\
\phantom{I_2}=&\int_{0}^{\infty} \! \! \frac{\d \omega'}{2\pi i}\, d_{\omega' k}  \theta \!\left( \omega^2-k^2\right) \theta\!\left( \omega'^2 - k^2\right) \frac{\delta\!\left(\sqrt{\omega'^2-k^2} - \sqrt{\omega^2 -k^2} \right)}{\sqrt{\omega^2-k^2}} \nn \\
\phantom{I_2}=&\int_{0}^{\infty} \! \! \frac{\d \omega'}{2\pi |\omega| i}\,d_{\omega' k}  \theta \!\left( \omega^2-k^2\right) \theta\!\left( \omega'^2 - k^2\right) \left( \delta \! \left( \omega - |\omega'|\right) + \delta\! \left( \omega+|\omega|\right) \right)\nn \\
\phantom{I_2}=&\theta \!\left( \omega^2 -k^2 \right) \frac{d_{|\omega| k}}{2\pi i |\omega|}
\end{align}
where we used  equations (\ref{appBessel 1}) and (\ref{deltafunction}).


The computation of $I_1$ goes as follows,
\begin{align}
&I_1=\int_0^\infty \!\!\! \d z \! \int_{-\infty}^\infty\!\! \! \d x \! \int_{-\infty}^{\infty} \! \! \frac{\d k'}{2\pi} \int_{-\infty}^{\infty} \! \! \frac{\d \omega'}{2\pi} \Bigg[ z \, \theta \! \left(\omega^2 - k^2\right)\frac{ \left(\omega'^2+k'^2\right)^{l/2}}{2^{l-1} \Gamma(l)}   \mathrm{e}^{i \omega' \epsilon - i (k-k') x}& \nn \\
&\phantom{I_1=\int_0^\infty \!\!\! \d z \! \int_{-\infty}^\infty\!\! \! \d x \! \int_{-\infty}^{\infty} \! \! \frac{\d k'}{2\pi} \int_{-\infty}^{\infty} \! \! \frac{\d \omega'}{2\pi}} J_l\left( \sqrt{ \omega^2 -k^2} \, z\right) K_l\left(\sqrt{\omega'^2+k'^2} \, z\right)\Bigg]& \nn \\
&\phantom{I_2}=\int_0^\infty \!\!\! \d z \! \int_{-\infty}^{\infty} \! \! \frac{\d \omega'}{2\pi} \Bigg[ z \, \theta \! \left(\omega^2 - k^2\right)\frac{ \left(\omega'^2+k^2\right)^{l/2}}{2^{l-1} \Gamma(l)}   \mathrm{e}^{i \omega' \epsilon} J_l\left( \sqrt{ \omega^2 -k^2} \, z\right) & \nn \\
&\phantom{I_2 =\int_0^\infty \!\!\! \d z \! \int_{-\infty}^{\infty} \! \! \frac{\d \omega'}{2\pi}} K_l\left(\sqrt{\omega'^2+k^2} \, z\right) \Bigg] \nn & \\
&\phantom{I_2}=\int_{-\infty}^{\infty} \! \! \frac{\d \omega'}{2\pi}  \frac{\theta \!\left(\omega^2 - k^2\right)}{2^{l-1} \Gamma(l)}   \mathrm{e}^{i \omega' \epsilon} \frac{\phantom{^{l}} \left(\omega^2 - k^2\right)^{l/2}}{\omega'^2 + \omega^2}& 
\end{align}
where for the last line we used equation (\ref{appBessel 2}).
The integral over $\omega'$ is performed using contour integration. 
Closing the contour in the upper half plane and picking up the contribution from the pole at $i|\omega|$ we obtain,
\begin{align}
I_1=& \frac{\theta \! \left( \omega^2 - k^2\right)\left(\omega^2 - k^2\right)^{l/2} }{2^l \pi \Gamma (l)} 2\pi i \, \text{Res}\left[\frac{\mathrm{e}^{i\omega' \epsilon}}{\omega'^2 + \omega^2} ; \omega' =i |\omega| \right] \nn \\
=& i \frac{\theta \!\left(\omega^2 - k^2 \right) \left(\omega^2 - k^2 \right)^{l/2} }{2^{l-1} \Gamma(l)} \left[\frac{\mathrm{e}^{-|\omega| \epsilon} }{2 i |\omega|}\right]
= \frac{ \theta \!\left( \omega^2 -k^2 \right) \left(\omega^2 - k^2 \right)^{l/2}}{2^l \Gamma(l) |\omega|} \mathrm{e}^{-|\omega| \epsilon}. 
\end{align}
Combining the results for $I_1$ and $I_2$,
\begin{align}
&\int_0^\infty \!\!\! \d z \, \theta\! \left(\omega^2 - k^2\right) J_l(\sqrt{\omega^2-k^2} \, z) \int_{-\infty}^{\infty}\!\! \d x \, \mathrm{e}^{-i kx} \left( \partial_{\tau_0} \Phi _E^- \left(\tau_0,z,x \right)\big|_{\tau_0=0}\right)=& \nonumber \\
&=\theta\!\left(\omega^2-k^2\right)\left(-\frac{\left(\omega^2 - k^2\right)^{l/2}}{2^l \Gamma(l)} \mathrm{e}^{-|\omega| \epsilon} + \frac{d_{|\omega| k} }{2\pi i} \right).&
\end{align}
The computations for the derivative follow along the same lines.

\bibliographystyle{JHEP-2}

\end{document}